\documentclass[oneside,finnish,english]{amsart}
\usepackage[T1]{fontenc}
\usepackage{textcomp}
\usepackage[utf8]{inputenc}
\usepackage{babel}
\usepackage{cprotect}
\usepackage{amstext}
\usepackage{amsthm}
\usepackage{amssymb}
\usepackage{graphicx}
\usepackage[unicode=true,
 bookmarks=true,bookmarksnumbered=false,bookmarksopen=false,
 breaklinks=false,pdfborder={0 0 1},backref=false,colorlinks=false]
 {hyperref}
\hypersetup{pdftitle={Multiple-scattering T-matrix approach in nanophotonics},
 pdfauthor={Marek Nečada}}

\makeatletter
\numberwithin{equation}{section}
\numberwithin{figure}{section}

\DeclareUnicodeCharacter{0428}{Ш }
\usepackage{tikz}
\usepackage{amsaddr}

\makeatother

\begin{document}
\title[$T$-matrix simulations: symmetries and periodic lattices]{Multiple-scattering $T$-matrix simulations for nanophotonics: symmetries
and periodic lattices}
\author{Marek Nečada and Päivi Törmä}
\address{Department of Applied Physics, Aalto University School of Science,
P.O. Box 15100, FI-00076 Aalto, Finland}
\email{marek@necada.org, paivi.torma@aalto.fi}
\begin{abstract}
The multiple scattering method T-matrix (MSTMM) can be used to solve
the electromagnetic response of systems consisting of many compact
scatterers, retaining a good level of accuracy while using relatively
few degrees of freedom, largely surpassing other methods in the number
of scatterers it can deal with. Here we extend the method to infinite
periodic structures using Ewald-type lattice summation, and we exploit
the possible symmetries of the structure to further improve its efficiency,
so that systems containing tens of thousands of particles can be studied
with relative ease. We release a modern implementation of the method,
including the theoretical improvements presented here, under GNU General
Public Licence.
\end{abstract}

\maketitle

\section*{}

\global\long\def\SI#1#2{}%
\global\long\def\uoft#1{\mathfrak{F}#1}%
\global\long\def\uaft#1{\mathfrak{\mathbb{F}}#1}%
\global\long\def\usht#1#2{\mathbb{S}_{#1}#2}%
\global\long\def\bsht#1#2{\mathrm{S}_{#1}#2}%
\global\long\def\sgn{\operatorname{sgn}}%
\global\long\def\pht#1#2{\mathfrak{\mathbb{H}}_{#1}#2}%
\global\long\def\vect#1{\mathbf{#1}}%
\global\long\def\uvec#1{\mathbf{\hat{#1}}}%
\global\long\def\ud{\mathrm{d}}%
\global\long\def\basis#1{\mathfrak{#1}}%
\global\long\def\dc#1{\mathrm{III}_{#1}}%
\global\long\def\rec#1{#1^{-1}}%
\global\long\def\recb#1{#1^{\widehat{-1}}}%
\global\long\def\ints{\mathbb{Z}}%
\global\long\def\nats{\mathbb{N}}%
\global\long\def\reals{\mathbb{R}}%
\global\long\def\ush#1#2{Y_{#1,#2}}%
\global\long\def\ushD#1#2{Y'_{#1,#2}}%
\global\long\def\vsh#1#2#3{\vect A_{#1,#2,#3}}%
\global\long\def\vshD#1#2#3{\vect A'_{#1,#2,#3}}%
\global\long\def\hgfr{\mathbf{F}}%
\global\long\def\hgf{F}%
\global\long\def\ghgf#1#2{\mbox{}_{#1}F_{#2}}%
\global\long\def\ghgfr#1#2{\mbox{}_{#1}\mathbf{F}_{#2}}%
\global\long\def\ph{\mathrm{ph}}%
\global\long\def\kor#1{\underline{#1}}%
\global\long\def\koru#1{\utilde{#1}}%
\global\long\def\swv{\mathscr{H}}%
\global\long\def\expint{\mathrm{E}}%
\global\long\def\thespace{\reals^{3}}%
\global\long\def\medium{\Theta}%
\global\long\def\mezikuli#1#2#3{\Theta_{#1,#2}\left(#3\right)}%
\global\long\def\epsbg{\mathrm{\epsilon_{b}}}%
\global\long\def\mubg{\mathrm{\mu_{b}}}%
\global\long\def\rcoeffp#1{a_{#1}}%
\global\long\def\rcoeffincp#1{\tilde{a}_{#1}}%
\global\long\def\rcoeff{a}%
\global\long\def\rcoeffinc{\tilde{a}}%
\global\long\def\rcoeffptlm#1#2#3#4{\rcoeffp{#1,#2#3#4}}%
\global\long\def\rcoefftlm#1#2#3{\rcoeffp{#1#2#3}}%
\global\long\def\rcoeffincptlm#1#2#3#4{\rcoeffincp{#1,#2#3#4}}%
\global\long\def\vswfrtlm#1#2#3{\vect v_{#1#2#3}}%
\global\long\def\sswfoutlm#1#2{\psi_{#1,#2}}%
\global\long\def\outcoeff{f}%
\global\long\def\outcoeffp#1{f_{#1}}%
\global\long\def\outcoeffptlm#1#2#3#4{\outcoeffp{#1,#2#3#4}}%
\global\long\def\outcoefftlm#1#2#3{\outcoeffp{#1#2#3}}%
\global\long\def\vswfouttlm#1#2#3{\vect u_{#1#2#3}}%
\global\long\def\Tp#1{T_{#1}}%
\global\long\def\openball#1#2{B_{#1}\left(#2\right)}%
\global\long\def\closedball#1#2{\overline{B_{#1}\left(#2\right)}}%
\global\long\def\tropr{\mathcal{R}}%
\global\long\def\troprp#1#2{\mathcal{\tropr}_{#1\leftarrow#2}}%
\global\long\def\trops{\mathcal{S}}%
\global\long\def\tropsp#1#2{\mathcal{\trops}_{#1\leftarrow#2}}%
\global\long\def\tropcoeff{C}%
\global\long\def\truncated#1#2{\left[#1\right]_{l\le#2}}%
\global\long\def\truncate#1#2{\left[#1\right]_{#2}}%
\global\long\def\dlmfFer#1#2{\mathsf{P}_{#1}^{#2}}%
\global\long\def\antidelta{\gamma}%
\global\long\def\groupop#1{\hat{P}_{#1}}%

\global\long\def\vswfr#1#2#3{\vswfrtlm{#3}{#1}{#2}}%
\global\long\def\vswfs#1#2#3{\vswfouttlm{#3}{#1}{#2}}%
\global\long\def\svwfs#1#2#3{\vswfouttlm{#3}{#1}{#2}}%
\global\long\def\coeffrip#1#2#3#4{\rcoeffptlm{#1}{#4}{#2}{#3}}%
\global\long\def\coeffsip#1#2#3#4{\outcoeffptlm{#1}{#4}{#2}{#3}}%
\global\long\def\coeffr{\rcoeffp{}}%
\global\long\def\coeffs{\outcoeffp{}}%
\global\long\def\transop{\trops}%
\global\long\def\coeffripext#1#2#3#4{\rcoeffincptlm{#1}{#4}{#2}{#3}}%
\global\long\def\Kp{K}%

\section{Introduction}\label{sec:Introduction}

The problem of electromagnetic response of a system consisting of
many relatively small, compact scatterers in various geometries, and
its numerical solution, is relevant to several branches of nanophotonics.
In practice, the scatterers often form some ordered structure, such
as metalic or dielectric nanoparticle arrays \cite{zou_silver_2004,garcia_de_abajo_colloquium:_2007,wang_rich_2018,kravets_plasmonic_2018}
that offer many degrees of tunability, with applications including
structural color, ultra-thin lenses \cite{khorasaninejad_metalenses_2017},
strong coupling between light and quantum emitters \cite{vakevainen_plasmonic_2014,ramezani_strong_2019,torma_strong_2015},
weak and strong coupling lasing and Bose-Einstein condensation \cite{zhou_lasing_2013,hakala_lasing_2017,guo_lasing_2019,hakala_boseeinstein_2018,yang_real-time_2015,ramezani_plasmon-exciton-polariton_2017,vakevainen_sub-picosecond_2020,wang_structural_2018},
magneto-optical effects \cite{kataja_surface_2015}, or sensing \cite{kuttner_plasmonics_2018}.
The number of scatterers tends to be rather large; unfortunately,
the most common general approaches used in computational electrodynamics
are often unsuitable for simulating systems with larger number of
scatterers due to their computational complexity: differential methods
such as the finite difference time domain (FDTD, \cite{sullivan_electromagnetic_2013})
method or the finite element method (FEM, \cite{raiyan_kabir_finite_2017})
include the field degrees of freedom (DoF) of the background medium
(which can have very large volumes), whereas integral approaches such
as the boundary element method (BEM, a.k.a the method of moments,
MOM \cite{harrington_field_1993,medgyesi-mitschang_generalized_1994,reid_efficient_2015})
need much less DoF but require working with dense matrices containing
couplings between each pair of DoF. Therefore, a common (frequency-domain)
approach to get an approximate solution of the scattering problem
for many small particles has been the coupled dipole approximation
(CD) \cite{zhao_extinction_2003} where a drastic reduction of the
number of DoF is achieved by approximating individual scatterers to
electric dipoles (characterised by a polarisability tensor) coupled
to each other through Green's functions. 

CD is easy to implement and demands relatively little computational
resources but suffers from at least two fundamental drawbacks. The
obvious one is that the dipole approximation is too rough for particles
with diameter larger than a small fraction of the wavelength, which
results to quantitative errors. The other one, more subtle, manifests
itself in photonic crystal-like structures used in nanophotonics:
there are modes in which the particles' electric dipole moments completely
vanish due to symmetry, and regardless of how small the particles
are, the excitations have quadrupolar or higher-degree multipolar
character. These modes, belonging to a category that is sometimes
called \emph{optical bound states in the continuum (BIC)} \cite{hsu_bound_2016},
typically appear at the band edges where interesting phenomena such
as lasing or Bose-Einstein condensation have been observed \cite{guo_lasing_2019,pourjamal_lasing_2019,hakala_lasing_2017,yang_real-time_2015,hakala_boseeinstein_2018}
-- and CD by definition fails to capture such modes.

The natural way to overcome both limitations of CD mentioned above
is to take higher multipoles into account. Instead of a polarisability
tensor, the scattering properties of an individual particle are then
described with more general \emph{transition matrix} (commonly known
as $T$-matrix), and different particles' multipole excitations are
coupled together via translation operators, a generalisation of the
Green's functions used in CDA. This is the idea behind the \emph{multiple-scattering
$T$-matrix method }(MSTMM), a.k.a. \emph{superposition $T$-matrix
method} \cite{litvinov_rigorous_2008}, and it has been implemented
many times in the context of electromagnetics \cite{scattport_multiple_nodate},
but usually only as specific codes for limited subsets of problems,
such as scattering by clusters of spheres, circular cylinders, or
Chebyshev particles \cite{mackowski_multiple_2011,mackowski_mstm_2013,xu_fortran_2003};
there also exists a code for modeling photonic slabs including 2D-periodic
infinite arrays of spheres \cite{stefanou_heterostructures_1998,stefanou_multem_2000}.
Perhaps the most general MSTMM software with respect to the system
geometry has been FaSTMM \cite{markkanen_fast_2017,markkanen_fastmm_2018},
which also a rare example is in this field of a publicly available
code with a clear licence.

However, the potential of MSTMM reaches far beyond its past implementations.
Here we present several enhancements to the method, which are especially
useful in metamaterial and nanophotonics simulations. We extend the
method on infinite periodic lattices (in all three possible dimensionalities)
using Ewald-type summation techniques. This enables, among other things,
to use MSTMM for fast solving of the lattice modes of such periodic
systems, and comparing them to their finite counterparts with respect
to electromagnetic response, which is useful to isolate the bulk and
finite-size phenomena of photonic lattices. Moreover, we exploit symmetries
of the system to decompose the problem into several substantially
smaller ones, which provides better understanding of modes, mainly
in periodic systems, and substantially reduces the demands on computational
resources, hence speeding up the computations and allowing for finite
size simulations of systems with particle numbers practically impossible
to reliably simulate with any other method. Furthemore, the method
can be combined with other integral methods, which removes the limitation
to systems with compact scatterers only, and enables e.g. including
a substrate \cite{czajkowski_multipole_2020}.

The power of the method has been already demonstrated by several works
where we used it to explain experimental observations: the finite
lattice size effects on dipole patterns and phase profiles of the
nanoparticle lattice modes in \cite{hakala_lasing_2017}, symmetry
and polarisation analysis of the modes at the $K$-point of a honeycomb
nanopatricle lattice in \cite{guo_lasing_2019}, the structure of
lasing modes in a Ni nanoparticle array \cite{pourjamal_lasing_2019}
and energy spacing between the $\Gamma$-point modes in a finite lattice
\cite{vakevainen_sub-picosecond_2020}. 

We hereby release our MSTMM implementation, the \emph{QPMS Photonic
Multiple Scattering} suite \cite{QPMS}, as free software under the
GNU General Public License version 3. QPMS allows for linear optics
simulations of arbitrary sets of compact scatterers in isotropic media.
The features include computations of electromagnetic response to external
driving, the related cross sections, and finding resonances of finite
structures. Moreover, it includes the improvements covered in this
article, enabling to simulate even larger systems and also infinite
structures with periodicity in one, two or three dimensions, which
can be used e.g. for evaluating dispersions of such structures. The
QPMS suite contains a core C library, Python bindings and several
utilities for routine computations, such as scattering cross sections
under plane wave irradiation or lattice modes of two-dimensional periodic
arrays. It includes Doxygen documentation together with description
of the API. It has been written with customisability and extendibility
in mind, so that including e.g. alternative methods of $T$-matrix
calculations of a single particle's matrix are as easy as possible.

The current paper is organised as follows: Section \ref{sec:Finite}
provides a review of MSTMM theory for finite systems. In Section \ref{sec:Infinite}
we develop the theory for infinite periodic structures. In Section
\ref{sec:Symmetries} we apply group theory on MSTMM to utilise the
symmetries of the simulated system. Finally, Section \ref{sec:Applications}
shows some practical results that can be obtained using QPMS. 

\section{Finite systems}\label{sec:Finite}

The basic idea of MSTMM is quite simple: the driving electromagnetic
field incident onto a scatterer is expanded into a vector spherical
wavefunction (VSWF) basis in which the single scattering problem is
solved, and the scattered field is then re-expanded into VSWFs centered
at the other scatterers. Repeating the same procedure with all (pairs
of) scatterers yields a set of linear equations, solution of which
gives the coefficients of the scattered field in the VSWF bases. Once
these coefficients have been found, one can evaluate various quantities
related to the scattering (such as cross sections or the scattered
fields) quite easily. 

The expressions appearing in the re-expansions are fairly complicated,
and the implementation of MSTMM is extremely error-prone also due
to the various conventions used in the literature. Therefore although
we do not re-derive from scratch the expressions that can be found
elsewhere in literature, for reader's reference we always state them
explicitly in our convention.

\subsection{Single-particle scattering}

In order to define the basic concepts, let us first consider the case
of electromagnetic (EM) radiation scattered by a single particle.
We assume that the scatterer lies inside a closed ball $\closedball{R^{<}}{\vect 0}$
of radius $R^{<}$ and center in the origin of the coordinate system
(which can be chosen that way; the natural choice of $\closedball{R^{<}}{\vect 0}$
is the circumscribed ball of the scatterer) and that there exists
a larger open cocentric ball $\openball{R^{>}}{\vect 0}$, such that
the (non-empty) spherical shell $\mezikuli{R^{<}}{R^{>}}{\vect 0}=\openball{R^{>}}{\vect 0}\setminus\closedball{R^{<}}{\vect 0}$
is filled with a homogeneous isotropic medium with relative electric
permittivity $\epsilon(\vect r,\omega)=\epsbg(\omega)$ and magnetic
permeability $\mu(\vect r,\omega)=\mubg(\omega)$, and that the whole
system is linear, i.e. the material properties of neither the medium
nor the scatterer depend on field intensities. Under these assumptions,
the EM fields $\vect{\Psi}=\vect E,\vect H$ in $\mezikuli{R^{<}}{R^{>}}{\vect 0}$
must satisfy the homogeneous vector Helmholtz equation together with
the transversality condition 
\begin{equation}
\left(\nabla^{2}+\kappa^{2}\right)\Psi\left(\vect r,\vect{\omega}\right)=0,\quad\nabla\cdot\vect{\Psi}\left(\vect r,\vect{\omega}\right)=0\label{eq:Helmholtz eq}
\end{equation}
   with wave number\footnote{Throughout this text, we use the letter $\kappa$ for wave number
in order to avoid confusion with Bloch vector $\vect k$ and its magnitude,
introduced in Section \ref{sec:Infinite}.} $\kappa=\kappa\left(\omega\right)=\omega\sqrt{\mubg(\omega)\epsbg(\omega)}/c_{0}$,
as can be derived from Maxwell's equations \cite{jackson_classical_1998}. 

\subsubsection{Spherical waves}

Equation \ref{eq:Helmholtz eq} can be solved by separation of variables
in spherical coordinates to give the solutions -- the \emph{regular}
and \emph{outgoing} vector spherical wavefunctions (VSWFs) $\vswfrtlm{\tau}lm\left(\kappa\vect r\right)$
and $\vswfouttlm{\tau}lm\left(\kappa\vect r\right)$, respectively,
defined as follows:
\begin{align}
\vswfrtlm 1lm\left(\kappa\vect r\right) & =j_{l}\left(\kappa r\right)\vsh 1lm\left(\uvec r\right),\nonumber \\
\vswfrtlm 2lm\left(\kappa\vect r\right) & =\frac{1}{\kappa r}\frac{\ud\left(\kappa rj_{l}\left(\kappa r\right)\right)}{\ud\left(\kappa r\right)}\vsh 2lm\left(\uvec r\right)+\sqrt{l\left(l+1\right)}\frac{j_{l}\left(\kappa r\right)}{\kappa r}\vsh 3lm\left(\uvec r\right),\label{eq:VSWF regular}
\end{align}
\begin{align}
\vswfouttlm 1lm\left(\kappa\vect r\right) & =h_{l}^{\left(1\right)}\left(\kappa r\right)\vsh 1lm\left(\uvec r\right),\nonumber \\
\vswfouttlm 2lm\left(\kappa\vect r\right) & =\frac{1}{kr}\frac{\ud\left(\kappa rh_{l}^{\left(1\right)}\left(\kappa r\right)\right)}{\ud\left(\kappa r\right)}\vsh 2lm\left(\uvec r\right)+\sqrt{l\left(l+1\right)}\frac{h_{l}^{\left(1\right)}\left(\kappa r\right)}{\kappa r}\vsh 3lm\left(\uvec r\right),\label{eq:VSWF outgoing}\\
 & \tau=1,2;\quad l=1,2,3,\dots;\quad m=-l,-l+1,\dots,+l,\nonumber 
\end{align}
where $\vect r=r\uvec r=r\left(\sin\theta\left(\uvec x\cos\phi+\uvec y\sin\phi\right)+\uvec z\cos\theta\right)$;
$j_{l}\left(x\right),h_{l}^{\left(1\right)}\left(x\right)=j_{l}\left(x\right)+iy_{l}\left(x\right)$
are the regular spherical Bessel function and spherical Hankel function
of the first kind\footnote{The interpretation of $\vswfouttlm{\tau}lm\left(\kappa\vect r\right)$
containing spherical Hankel functions of the first kind as \emph{outgoing}
waves at positive frequencies is associated with a specific choice
of sign in the exponent of time-frequency transformation, $\psi\left(t\right)=\left(2\pi\right)^{-1/2}\int\psi\left(\omega\right)e^{-i\omega t}\,\ud\omega$.
This matters especially when considering materials with gain or loss:
in this convention, lossy materials will have refractive index (and
wavenumber $\kappa$, at a given positive frequency) with \emph{positive}
imaginary part, and gain materials will have it negative and, for
example, Drude-Lorenz model of a lossy medium will have poles in the
lower complex half-plane.}, respectively, as in \cite[§10.47]{NIST:DLMF}, and $\vsh{\tau}lm$
are the \emph{vector spherical harmonics}
\begin{align}
\vsh 1lm\left(\uvec r\right) & =\frac{1}{\sqrt{l\left(l+1\right)}}\nabla\times\left(\vect r\ush lm\left(\uvec r\right)\right)=\frac{1}{\sqrt{l\left(l+1\right)}}\nabla\ush lm\left(\uvec r\right)\times\vect r,\nonumber \\
\vsh 2lm\left(\uvec r\right) & =\frac{1}{\sqrt{l\left(l+1\right)}}r\nabla\ush lm\left(\uvec r\right),\nonumber \\
\vsh 3lm\left(\uvec r\right) & =\uvec r\ush lm\left(\uvec r\right).\label{eq:vector spherical harmonics definition}
\end{align}
Note that the regular waves $\vswfrtlm{\tau}lm$ (with fields expressed
in cartesian coordinates) have all well-defined limits in the origin,
and except for the ``electric dipolar'' waves $\vswfrtlm 21m$,
they vanish. In our convention, the (scalar) spherical harmonics $\ush lm$
are identical to those in \cite[14.30.1]{NIST:DLMF}, i.e.
\[
\ush lm\left(\uvec r\right)=\left(\frac{\left(l-m\right)!\left(2l+1\right)}{4\pi\left(l+m\right)!}\right)^{\frac{1}{2}}e^{im\phi}\dlmfFer lm\left(\cos\theta\right)
\]
where $ $ importantly, the Ferrers functions $\dlmfFer lm$ defined
as in \cite[§14.3(i)]{NIST:DLMF} do already contain the Condon-Shortley
phase $\left(-1\right)^{m}$. For later use, we also introduce ``dual''
spherical harmonics $\ushD lm$ defined by duality relation with the
``usual'' spherical harmonics 
\begin{equation}
\iint\ushD{l'}{m'}\left(\uvec r\right)\ush lm\left(\uvec r\right)\,\ud\Omega=\delta_{\tau'\tau}\delta_{l'l}\delta_{m'm}\label{eq:dual spherical harmonics}
\end{equation}
 and corresponding dual vector spherical harmonics 
\begin{equation}
\iint\vshD{\tau'}{l'}{m'}\left(\uvec r\right)\cdot\vsh{\tau}lm\left(\uvec r\right)\,\ud\Omega=\delta_{\tau'\tau}\delta_{l'l}\delta_{m'm}\label{eq:dual vsh}
\end{equation}
(complex conjugation not implied in the dot product here). In our
convention, we have
\begin{align*}
\ush lm\left(\uvec r\right) & =\left(\ush lm\left(\uvec r\right)\right)^{*}=\left(-1\right)^{m}\ush l{-m}\left(\uvec r\right).\\
\vshD{\tau}lm\left(\uvec r\right) & =\left(\vsh{\tau}lm\left(\uvec r\right)\right)^{*}=\left(-1\right)^{m}\vsh{\tau}l{-m}\left(\uvec r\right).
\end{align*}

The convention for VSWFs used here is the same as in \cite{kristensson_spherical_2014};
over other conventions used elsewhere in literature, it has several
fundamental advantages -- most importantly, the translation operators
introduced later in eq. (\ref{eq:reqular vswf coefficient vector translation})
are unitary, and it gives the simplest possible expressions for power
transport and cross sections without additional $l,m$-dependent factors
(for that reason, we also call our VSWFs as \emph{power-normalised}).
Power-normalisation and unitary translation operators are possible
to achieve also with real spherical harmonics -- such a convention
is used in \cite{kristensson_scattering_2016}.

\subsubsection{T-matrix definition}

The regular VSWFs $\vswfrtlm{\tau}lm\left(k\vect r\right)$ would
constitute a basis for solutions of the Helmholtz equation (\ref{eq:Helmholtz eq})
inside a ball $\openball{R^{>}}{\vect 0}$ with radius $R^{>}$ and
center in the origin, were it filled with homogeneous isotropic medium;
however, if the equation is not guaranteed to hold inside a smaller
ball $\closedball{R^{<}}{\vect 0}$ around the origin (typically due
to presence of a scatterer), one has to add the outgoing VSWFs $\vswfouttlm{\tau}lm\left(\kappa\vect r\right)$
to have a complete basis of the solutions in the volume $\mezikuli{R^{<}}{R^{>}}{\vect 0}=\openball{R^{>}}{\vect 0}\setminus\closedball{R^{<}}{\vect 0}$.

The single-particle scattering problem at frequency $\omega$ can
be posed as follows: Let a scatterer be enclosed inside the ball $\closedball{R^{<}}{\vect 0}$
and let the whole volume $\mezikuli{R^{<}}{R^{>}}{\vect 0}$ be filled
with a homogeneous isotropic medium with wave number $\kappa\left(\omega\right)$.
Inside $\mezikuli{R^{<}}{R^{>}}{\vect 0}$, the electric field can
be expanded as
\begin{equation}
\vect E\left(\omega,\vect r\right)=\sum_{\tau=1,2}\sum_{l=1}^{\infty}\sum_{m=-l}^{+l}\left(\rcoefftlm{\tau}lm\vswfrtlm{\tau}lm\left(\kappa\vect r\right)+\outcoefftlm{\tau}lm\vswfouttlm{\tau}lm\left(\kappa\vect r\right)\right).\label{eq:E field expansion}
\end{equation}
If there were no scatterer and $\closedball{R^{<}}{\vect 0}$ were
filled with the same homogeneous medium, the part with the outgoing
VSWFs would vanish and only the part $\vect E_{\mathrm{inc}}=\sum_{\tau lm}\rcoefftlm{\tau}lm\vswfrtlm{\tau}lm$
due to sources outside $\openball{R^{>}}{\vect 0}$ would remain.
Let us assume that the ``driving field'' is given, so that presence
of the scatterer does not affect $\vect E_{\mathrm{inc}}$ and is
fully manifested in the latter part, $\vect E_{\mathrm{scat}}=\sum_{\tau lm}\outcoefftlm{\tau}lm\vswfouttlm{\tau}lm$.
We also assume that the scatterer is made of optically linear materials
and hence reacts to the incident field in a linear manner. This gives
a linearity constraint between the expansion coefficients
\begin{equation}
\outcoefftlm{\tau}lm=\sum_{\tau'l'm'}T_{\tau lm;\tau'l'm'}\rcoefftlm{\tau'}{l'}{m'}\label{eq:T-matrix definition}
\end{equation}
where the $T_{\tau lm;\tau'l'm'}=T_{\tau lm;\tau'l'm'}\left(\omega\right)$
are the elements of the \emph{transition matrix,} a.k.a. $T$-matrix.
It completely describes the scattering properties of a linear scatterer,
so with the knowledge of the $T$-matrix we can solve the single-particle
scattering problem simply by substituting appropriate expansion coefficients
$\rcoefftlm{\tau'}{l'}{m'}$ of the driving field into (\ref{eq:T-matrix definition}).
The outgoing VSWF expansion coefficients $\outcoefftlm{\tau}lm$ are
the effective induced electric ($\tau=2$) and magnetic ($\tau=1$)
multipole polarisation amplitudes of the scatterer, and this is why
we sometimes refer to the corresponding VSWFs as to the electric and
magnetic VSWFs, respectively. 

$T$-matrices of particles with certain simple geometries (most famously
spherical) can be obtained analytically \cite{kristensson_scattering_2016,mie_beitrage_1908};
for particles with smooth surfaces one can find them numerically using
the \emph{null-field method} \cite{waterman_new_1969,waterman_symmetry_1971,kristensson_scattering_2016}
which works well in the most typical cases, but for less common parameter
ranges (such as concave shapes, extreme values of aspect ratios or
relative refractive index) they might suffer from serious numerical
instabilities \cite[Sect. 5.8.4]{mishchenko_scattering_2002}. In
general, elements of the $T$-matrix can be obtained by simulating
scattering of a regular spherical wave $\vswfrtlm{\tau}lm$ and projecting
the scattered fields (or induced currents, depending on the method)
onto the outgoing VSWFs $\vswfouttlm{\tau'}{l'}{m'}$. In practice,
one can compute only a finite number of elements with a cut-off value
$L$ on the multipole degree, $l,l'\le L$, see below. 

For the numerical evaluation of $T$-matrices for simple axially symmetric
scatterers in QPMS, we typically use the null-field equations, and
for more complicated scatterers we use the \texttt{scuff-tmatrix}
tool from the free software SCUFF-EM suite \cite{reid_efficient_2015,SCUFF2}.\footnote{Note that the upstream versions of SCUFF-EM contain a bug that renders
almost all $T$-matrix results wrong; we found and fixed the bug in
our fork available at \href{https://github.com/texnokrates/scuff-em}{https://github.com/texnokrates/scuff-em}
in revision \href{https://github.com/texnokrates/scuff-em/commit/78689f5514072853aa5cad455ce15b3e024d163d}{g78689f5}.
However, the \href{https://github.com/HomerReid/scuff-em/pull/197}{bugfix}
has not been merged into upstream by the time of writing this article. }

\subsubsection{T-matrix compactness, cutoff validity}

The magnitude of the $T$-matrix elements depends heavily on the scatterer's
size compared to the wavelength. Typically,\footnote{It has been proven that the $T$-matrix of a bounded scatterer is
a compact operator for \emph{acoustic} scattering problems \cite{ganesh_convergence_2012}.
While we conjecture that this holds also for bounded electromagnetic
scatterers, we are not aware of a definitive proof.} from certain multipole degree onwards, $l,l'>L$, the elements of
the $T$-matrix are negligible, so truncating the $T$-matrix at finite
multipole degree $L$ gives a good approximation of the actual infinite-dimensional
operator. If the incident field is well-behaved, i.e. the expansion
coefficients $\rcoefftlm{\tau'}{l'}{m'}$ do not take excessive values
for $l'>L$, the scattered field expansion coefficients $\outcoefftlm{\tau}lm$
with $l>L$ will also be negligible.

A rule of thumb to choose the $L$ with desired $T$-matrix element
accuracy $\delta$ can be obtained from the spherical Bessel function
expansion around zero \cite[10.52.1]{NIST:DLMF} by requiring that
$\delta\gg\left(nR\right)^{L}/\left(2L+1\right)!!$, where $R$ is
the scatterer radius and $n$ its (maximum) refractive index.

\subsubsection{Power transport}

For convenience, let us introduce a short-hand matrix notation for
the expansion coefficients and related quantities, so that we do not
need to write the indices explicitly; so for example, eq. (\ref{eq:T-matrix definition})
would be written as $\outcoeffp{}=T\rcoeffp{}$, where $\rcoeffp{},\outcoeffp{}$
are column vectors with the expansion coefficients. Transposed and
complex-conjugated matrices are labeled with the $\dagger$ superscript.

With this notation, we state an important result about power transport,
derivation of which can be found in \cite[sect. 7.3]{kristensson_scattering_2016}.
Let the field in $\mezikuli{R^{<}}{R^{>}}{\vect 0}$ have expansion
as in (\ref{eq:E field expansion}). Then the net power transported
from $\openball{R^{<}}{\vect 0}$ to $\mezikuli{R^{<}}{R^{>}}{\vect 0}$
by electromagnetic radiation is
\begin{equation}
P=\frac{1}{2\kappa^{2}\eta_{0}\eta}\left(\Re\left(\rcoeffp{}^{\dagger}\outcoeffp{}\right)+\left\Vert \outcoeffp{}\right\Vert ^{2}\right)=\frac{1}{2\kappa^{2}\eta_{0}\eta}\rcoeffp{}^{\dagger}\left(\Tp{}^{\dagger}\Tp{}+\frac{\Tp{}^{\dagger}+\Tp{}}{2}\right)\rcoeffp{},\label{eq:Power transport}
\end{equation}
where $\eta_{0}=\sqrt{\mu_{0}/\varepsilon_{0}}$ and $\eta=\sqrt{\mu/\varepsilon}$
are wave impedance of vacuum and relative wave impedance of the medium
in $\mezikuli{R^{<}}{R^{>}}{\vect 0}$, respectively. Here $P$ is
well-defined only when $\kappa^{2}\eta$ is real. In realistic scattering
setups, power is transferred by radiation into $\openball{R^{<}}{\vect 0}$
and absorbed by the enclosed scatterer, so $P$ is negative and its
magnitude equals to power absorbed by the scatterer. In other words,
the hermitian operator $\Pi=\Tp{}^{\dagger}\Tp{}+\left(\Tp{}^{\dagger}+\Tp{}\right)/2$
must be negative (semi-)definite for a particle without gain. This
provides a simple but very useful sanity check on the numerically
obtained $T$-matrices: non-negligible positive eigenvalues of $\Pi$
indicate either too drastic multipole truncation or another problem
with the $T$-matrix.

\subsubsection{Plane wave expansion}

In many scattering problems considered in practice, the driving field
is at least approximately a plane wave. A transversal ($\uvec k\cdot\vect E_{0}=0$)
plane wave propagating in direction $\uvec k$ with (complex) amplitude
$\vect E_{0}$ can be expanded into regular VSWFs \cite[7.7.1]{kristensson_scattering_2016}
as
\[
\vect E_{\mathrm{PW}}\left(\vect r,\omega\right)=\vect E_{0}e^{i\kappa\uvec k\cdot\vect r}=\sum_{\tau,l,m}\rcoeffptlm{}{\tau}lm\left(\uvec k,\vect E_{0}\right)\vswfrtlm{\tau}lm\left(\kappa\vect r\right),
\]
where the expansion coefficients are obtained from the scalar products
of the amplitude and corresponding dual vector spherical harmonics
\begin{eqnarray}
\rcoeffptlm{}1lm\left(\uvec k,\vect E_{0}\right) & = & 4\pi i^{l}\vshD 1lm\left(\uvec k\right)\cdot\vect E_{0},\nonumber \\
\rcoeffptlm{}2lm\left(\uvec k,\vect E_{0}\right) & = & -4\pi i^{l+1}\vshD 2lm\left(\uvec k\right)\cdot\vect E_{0}.\label{eq:plane wave expansion}
\end{eqnarray}

\subsubsection{Cross-sections (single-particle)}

With the $T$-matrix and expansion coefficients of plane waves in
hand, we can state the expressions for cross-sections of a single
scatterer. Assuming a non-lossy background medium, extinction, scattering
and absorption cross sections of a single scatterer irradiated by
a plane wave propagating in direction $\uvec k$ and (complex) amplitude
$\vect E_{0}$ are \cite[sect. 7.8.2]{kristensson_scattering_2016}
\begin{eqnarray}
\sigma_{\mathrm{ext}}\left(\uvec k\right) & = & -\frac{1}{\kappa^{2}\left\Vert \vect E_{0}\right\Vert ^{2}}\Re\left(\rcoeffp{}^{\dagger}\outcoeffp{}\right)=-\frac{1}{2\kappa^{2}\left\Vert \vect E_{0}\right\Vert ^{2}}\rcoeffp{}^{\dagger}\left(\Tp{}+\Tp{}^{\dagger}\right)\rcoeffp{},\label{eq:extincion CS single}\\
\sigma_{\mathrm{scat}}\left(\uvec k\right) & = & \frac{1}{\kappa^{2}\left\Vert \vect E_{0}\right\Vert ^{2}}\left\Vert \outcoeffp{}\right\Vert ^{2}=\frac{1}{\kappa^{2}\left\Vert \vect E_{0}\right\Vert ^{2}}\rcoeffp{}^{\dagger}\left(\Tp{}^{\dagger}\Tp{}\right)\rcoeffp{},\label{eq:scattering CS single}\\
\sigma_{\mathrm{abs}}\left(\uvec k\right) & = & \sigma_{\mathrm{ext}}\left(\uvec k\right)-\sigma_{\mathrm{scat}}\left(\uvec k\right)=-\frac{1}{\kappa^{2}\left\Vert \vect E_{0}\right\Vert ^{2}}\left(\Re\left(\rcoeffp{}^{\dagger}\outcoeffp{}\right)+\left\Vert \outcoeffp{}\right\Vert ^{2}\right)\nonumber \\
 &  & =-\frac{1}{\kappa^{2}\left\Vert \vect E_{0}\right\Vert ^{2}}\rcoeffp{}^{\dagger}\left(\Tp{}^{\dagger}\Tp{}+\frac{\Tp{}^{\dagger}+\Tp{}}{2}\right)\rcoeffp{},\label{eq:absorption CS single}
\end{eqnarray}
where $\rcoeffp{}=\rcoeffp{}\left(\vect k,\vect E_{0}\right)$ is
the vector of plane wave expansion coefficients as in (\ref{eq:plane wave expansion}). 

\subsection{Multiple scattering}\label{subsec:Multiple-scattering}

If the system consists of multiple scatterers, the EM fields around
each one can be expanded in analogous way. Let $\mathcal{P}$ be an
index set labeling the scatterers. We enclose each scatterer in a
closed ball $\closedball{R_{p}}{\vect r_{p}}$ such that the balls
do not touch, $\closedball{R_{p}}{\vect r_{p}}\cap\closedball{R_{q}}{\vect r_{q}}=\emptyset;p,q\in\mathcal{P}$,
so there is a non-empty spherical shell $\mezikuli{R_{p}}{R_{p}^{>}}{\vect r_{p}}$
around each one that contains only the background medium without any
scatterers; we assume that all the relevant volume outside $\bigcap_{p\in\mathcal{P}}\closedball{R_{p}}{\vect r_{p}}$
is filled with the same background medium. Then the EM field inside
each $\mezikuli{R_{p}}{R_{p}^{>}}{\vect r_{p}}$ can be expanded in
a way similar to (\ref{eq:E field expansion}), using VSWFs with origins
shifted to the centre of the volume:
\begin{align}
\vect E\left(\omega,\vect r\right) & =\sum_{\tau=1,2}\sum_{l=1}^{\infty}\sum_{m=-l}^{+l}\left(\rcoeffptlm p{\tau}lm\vswfrtlm{\tau}lm\left(\kappa\left(\vect r-\vect r_{p}\right)\right)+\outcoeffptlm p{\tau}lm\vswfouttlm{\tau}lm\left(\kappa\left(\vect r-\vect r_{p}\right)\right)\right),\label{eq:E field expansion multiparticle}\\
 & \vect r\in\mezikuli{R_{p}}{R_{p}^{>}}{\vect r_{p}}.\nonumber 
\end{align}
Unlike the single scatterer case, the incident field coefficients
$\rcoeffptlm p{\tau}lm$ here are not only due to some external driving
field that the particle does not influence but they also contain the
contributions of fields scattered from \emph{all other scatterers}:
\begin{equation}
\rcoeffp p=\rcoeffincp p+\sum_{q\in\mathcal{P}\backslash\left\{ p\right\} }\tropsp pq\outcoeffp q\label{eq:particle total incident field coefficient a}
\end{equation}
where $\rcoeffincp p$ represents the part due to the external driving
that the scatterers can not influence, and $\tropsp pq$ is a \emph{translation
operator} defined below in Sec. \ref{subsec:Translation-operator},
that contains the re-expansion coefficients of the outgoing waves
in origin $\vect r_{q}$ into regular waves in origin $\vect r_{p}$.
For each scatterer, we also have its $T$-matrix relation as in (\ref{eq:T-matrix definition}),
\[
\outcoeffp q=T_{q}\rcoeffp q.
\]
Together with (\ref{eq:particle total incident field coefficient a}),
this gives rise to a set of linear equations
\begin{equation}
\outcoeffp p-T_{p}\sum_{q\in\mathcal{P}\backslash\left\{ p\right\} }\tropsp pq\outcoeffp q=T_{p}\rcoeffincp p,\quad p\in\mathcal{P}\label{eq:Multiple-scattering problem}
\end{equation}
which defines the multiple-scattering problem. If all the $p,q$-indexed
vectors and matrices (note that without truncation, they are infinite-dimensional)
are arranged into blocks of even larger vectors and matrices, this
can be written in a short-hand form
\begin{equation}
\left(I-T\trops\right)\outcoeff=T\rcoeffinc\label{eq:Multiple-scattering problem block form}
\end{equation}
where $I$ is the identity matrix, $T$ is a block-diagonal matrix
containing all the individual $T$-matrices, and $\trops$ contains
the individual $\tropsp pq$matrices as the off-diagonal blocks, whereas
the diagonal blocks are set to zeros. 

We note that eq. (\ref{eq:Multiple-scattering problem block form})
with zero right-hand side describes the normal modes of the system;
the methods mentioned later in Section (\ref{sec:Infinite}) for solving
the band structure of a periodic system can be used as well for finding
the resonant frequencies of a finite system.

In practice, the multiple-scattering problem is solved in its truncated
form, in which all the $l$-indices related to a given scatterer $p$
are truncated as $l\le L_{p}$, leaving only $N_{p}=2L_{p}\left(L_{p}+2\right)$
different $\tau lm$-multi-indices left. The truncation degree can
vary for different scatterers (e.g.\ due to different physical sizes),
so the truncated block $\left[\tropsp pq\right]_{l_{q}\le L_{q};l_{p}\le L_{q}}$
has shape $N_{p}\times N_{q}$, not necessarily square. 

If no other type of truncation is done, there remain $2L_{p}\left(L_{p}+2\right)$
different $\tau lm$-multi-indices for the $p$-th scatterer, so that
the truncated version of the matrix $\left(I-T\trops\right)$ is a
square matrix with $\left(\sum_{p\in\mathcal{P}}N_{p}\right)^{2}$
elements in total. The truncated problem (\ref{eq:Multiple-scattering problem block form})
can then be solved using standard numerical linear algebra methods
(typically, by LU factorisation of the $\left(I-T\trops\right)$ matrix
at a given frequency, and then solving with Gauss elimination for
as many different incident $\rcoeffinc$ vectors as needed).

Alternatively, the multiple scattering problem can be formulated in
terms of the regular field expansion coefficients,
\begin{align*}
\rcoeffp p-\sum_{q\in\mathcal{P}\backslash\left\{ p\right\} }\tropsp pqT_{q}\rcoeffp q & =\rcoeffincp p,\quad p\in\mathcal{P},\\
\left(I-\trops T\right)\rcoeff & =\rcoeffinc,
\end{align*}
but this form is less suitable for numerical calculations due to the
fact that the regular VSWF expansion coefficients on both sides of
the equation are typically non-negligible even for large multipole
degree $l$, hence the truncation is not justified in this case. 

\subsubsection{Translation operator}\label{subsec:Translation-operator}

Let $\vect r_{1},\vect r_{2}$ be two different origins; a regular
VSWF with origin $\vect r_{1}$ can be expanded in terms of regular
VSWFs with origin $\vect r_{2}$ as follows:

\begin{equation}
\vswfrtlm{\tau}lm\left(\kappa\left(\vect r-\vect r_{1}\right)\right)=\sum_{\tau'l'm'}\tropr_{\tau lm;\tau'l'm'}\left(\kappa\left(\vect r_{2}-\vect r_{1}\right)\right)\vswfrtlm{\tau'}{l'}{m'}\left(\vect r-\vect r_{2}\right),\label{eq:regular vswf translation}
\end{equation}
where an explicit formula for the regular translation operator $\tropr$
reads in eq. (\ref{eq:translation operator}) below. For singular
(outgoing) waves, the form of the expansion differs inside and outside
the ball $\closedball{\left\Vert \vect r_{2}-\vect r_{1}\right\Vert }{\vect r_{1}}$:
\begin{multline}
\vswfouttlm{\tau}lm\left(\kappa\left(\vect r-\vect r_{1}\right)\right)=\\
=\begin{cases}
\sum_{\tau'l'm'}\trops_{\tau lm;\tau'l'm'}\left(\kappa\left(\vect r_{2}-\vect r_{1}\right)\right)\vswfrtlm{\tau'}{l'}{m'}\left(\kappa\left(\vect r-\vect r_{2}\right)\right), & \vect r\in\openball{\left\Vert \vect r_{1}-\vect r_{2}\right\Vert }{\vect r_{2}}\\
\sum_{\tau'l'm'}\tropr_{\tau lm;\tau'l'm'}\left(\kappa\left(\vect r_{2}-\vect r_{1}\right)\right)\vswfouttlm{\tau'}{l'}{m'}\left(\kappa\left(\vect r-\vect r_{2}\right)\right), & \vect r\notin\closedball{\left\Vert \vect r_{1}-\vect r_{2}\right\Vert }{\vect r_{2}}
\end{cases},\label{eq:singular vswf translation}
\end{multline}
where the singular translation operator $\trops$ has the same form
as $\tropr$ in (\ref{eq:translation operator}) except the regular
spherical Bessel functions $j_{l}$ are replaced with spherical Hankel
functions $h_{l}^{(1)}$.  

As MSTMM deals most of the time with the \emph{expansion coefficients}
of fields $\rcoeffptlm p{\tau}lm,\outcoeffptlm p{\tau}lm$ in different
origins $\vect r_{p}$ rather than with the VSWFs directly, let us
write down how \emph{they} transform under translation. We assume
the field can be expressed in terms of regular waves everywhere, and
expand it in two different origins $\vect r_{p},\vect r_{q}$,
\[
\vect E\left(\vect r,\omega\right)=\sum_{\tau,l,m}\rcoeffptlm p{\tau}lm\vswfrtlm{\tau}lm\left(\kappa\left(\vect r-\vect r_{p}\right)\right)=\sum_{\tau',l',m'}\rcoeffptlm q{\tau'}{l'}{m'}\vswfrtlm{\tau}{'l'}{m'}\left(\kappa\left(\vect r-\vect r_{q}\right)\right).
\]
Re-expanding the waves around $\vect r_{p}$ in terms of waves around
$\vect r_{q}$ using (\ref{eq:regular vswf translation}), 
\[
\vect E\left(\vect r,\omega\right)=\sum_{\tau,l,m}\rcoeffptlm p{\tau}lm\sum_{\tau'l'm'}\tropr_{\tau lm;\tau'l'm'}\left(\kappa\left(\vect r_{q}-\vect r_{p}\right)\right)\vswfrtlm{\tau'}{l'}{m'}\left(\kappa\left(\vect r-\vect r_{q}\right)\right)
\]
and comparing to the original expansion around $\vect r_{q}$, we
obtain
\begin{equation}
\rcoeffptlm q{\tau'}{l'}{m'}=\sum_{\tau,l,m}\tropr_{\tau lm;\tau'l'm'}\left(\kappa\left(\vect r_{q}-\vect r_{p}\right)\right)\rcoeffptlm p{\tau}lm.\label{eq:regular vswf coefficient translation}
\end{equation}
For the sake of readability, we introduce a shorthand matrix form
for (\ref{eq:regular vswf coefficient translation})
\begin{equation}
\rcoeffp q=\troprp qp\rcoeffp p\label{eq:reqular vswf coefficient vector translation}
\end{equation}
(note the reversed indices) Similarly, if we had only outgoing waves
in the original expansion around $\vect r_{p}$, we would get
\begin{equation}
\rcoeffp q=\tropsp qp\outcoeffp p\label{eq:singular to regular vswf coefficient vector translation}
\end{equation}
for the expansion inside the ball $\openball{\left\Vert \vect r_{q}-\vect r_{p}\right\Vert }{\vect r_{p}}$
 and 
\begin{equation}
\outcoeffp q=\troprp qp\outcoeffp p\label{eq:singular to singular vswf coefficient vector translation-1}
\end{equation}
outside.

In our convention, the regular translation operator elements can be
expressed explicitly as 
\begin{align}
\tropr_{\tau lm;\tau'l'm'}\left(\vect d\right) & =\sum_{\lambda=\left|l-l'\right|+\left|\tau-\tau'\right|}^{l+l'}\tropcoeff_{\tau lm;\tau'l'm'}^{\lambda}\ush{\lambda}{m-m'}\left(\uvec d\right)j_{\lambda}\left(d\right),\label{eq:translation operator}
\end{align}
and analogously the elements of the singular operator $\trops$, having
spherical Hankel functions ($h_{l}^{(1)}=j_{l}+iy_{l}$) in the radial
part instead of the regular Bessel functions,
\begin{align}
\trops_{\tau lm;\tau'l'm'}\left(\vect d\right) & =\sum_{\lambda=\left|l-l'\right|+\left|\tau-\tau'\right|}^{l+l'}\tropcoeff_{\tau lm;\tau'l'm'}^{\lambda}\ush{\lambda}{m-m'}\left(\uvec d\right)h_{\lambda}^{(1)}\left(d\right),\label{eq:translation operator singular}
\end{align}
where the constant factors in our convention read
\[
\tropcoeff_{\tau lm;\tau'l'm'}^{\lambda}=\begin{cases}
A_{lm;l'm'}^{\lambda} & \tau=\tau',\\
B_{lm;l'm'}^{\lambda} & \tau\ne\tau',
\end{cases}
\]
\begin{multline}
A_{lm;l'm'}^{\lambda}=\left(-1\right)^{\frac{l'-l+\lambda}{2}}\sqrt{\frac{4\pi\left(2\lambda+1\right)\left(2l+1\right)\left(2l'+1\right)}{l\left(l+1\right)l'\left(l'+1\right)}}\times\\
\times\begin{pmatrix}l & l' & \lambda\\
0 & 0 & 0
\end{pmatrix}\begin{pmatrix}l & l' & \lambda\\
m & -m' & m'-m
\end{pmatrix}\left(l\left(l+1\right)+l'\left(l'+1\right)-\lambda\left(\lambda+1\right)\right),\\
B_{lm;l'm'}^{\lambda}=-i\left(-1\right)^{\frac{l'-l+\lambda+1}{2}}\sqrt{\frac{4\pi\left(2\lambda+1\right)\left(2l+1\right)\left(2l'+1\right)}{l\left(l+1\right)l'\left(l'+1\right)}}\times\\
\times\begin{pmatrix}l & l' & \lambda-1\\
0 & 0 & 0
\end{pmatrix}\begin{pmatrix}l & l' & \lambda\\
m & -m' & m'-m
\end{pmatrix}\sqrt{\lambda^{2}-\left(l-l'\right)^{2}}\sqrt{\left(l+l'+1\right)^{2}-\lambda^{2}}.\label{eq:translation operator constant factors}
\end{multline}
Here $\begin{pmatrix}l_{1} & l_{2} & l_{3}\\
m_{1} & m_{2} & m_{3}
\end{pmatrix}$ is the $3j$ symbol defined as in \cite[§34.2]{NIST:DLMF}. Importantly
for practical calculations, these rather complicated coefficients
need to be evaluated only once up to the highest truncation order,
$l,l'\le L$. 

In our convention, the regular translation operator is unitary, $\left(\tropr_{\tau lm;\tau'l'm'}\left(\vect d\right)\right)^{-1}=\tropr_{\tau lm;\tau'l'm'}\left(-\vect d\right)=\tropr_{\tau'l'm';\tau lm}^{*}\left(\vect d\right)$,
or in the per-particle matrix notation, 
\begin{equation}
\troprp qp^{-1}=\troprp pq=\troprp qp^{\dagger}.\label{eq:regular translation unitarity}
\end{equation}
 Note that truncation at finite multipole degree breaks the unitarity,
$\truncated{\troprp qp}L^{-1}\ne\truncated{\troprp pq}L=\truncated{\troprp qp^{\dagger}}L$,
which has to be taken into consideration when evaluating quantities
such as absorption or scattering cross sections. Similarly, the full
regular operators can be composed
\begin{equation}
\troprp ac=\troprp ab\troprp bc\label{eq:regular translation composition}
\end{equation}
 but truncation breaks this, $\truncated{\troprp ac}L\ne\truncated{\troprp ab}L\truncated{\troprp bc}L.$

\subsubsection{Cross-sections (many scatterers)}

For a system of many scatterers, Kristensson \cite[sect. 9.2.2]{kristensson_scattering_2016}
derives only the extinction cross section formula. Let us re-derive
it together with the many-particle scattering and absorption cross
sections. First, let us take a ball containing all the scatterers
at once, $\openball R{\vect r_{\square}}\supset\bigcup_{p\in\mathcal{P}}\closedball{R_{p}}{\vect r_{p}}$.
Outside $\openball R{\vect r_{\square}}$, we can describe the EM
fields as if there was only a single scatterer,
\[
\vect E\left(\vect r\right)=\sum_{\tau,l,m}\left(\rcoeffptlm{\square}{\tau}lm\vswfrtlm{\tau}lm\left(\kappa\left(\vect r-\vect r_{\square}\right)\right)+\outcoeffptlm{\square}{\tau}lm\vswfouttlm{\tau}lm\left(\kappa\left(\vect r-\vect r_{\square}\right)\right)\right),
\]
where $\rcoeffp{\square},\outcoeffp{\square}$ are the vectors of
VSWF expansion coefficients of the incident and total scattered fields,
respectively, at origin $\vect r_{\square}$. In principle, one could
evaluate $\outcoeffp{\square}$ using the translation operators and
use the single-scatterer formulae (\ref{eq:extincion CS single})--(\ref{eq:absorption CS single})
with $\rcoeffp{}=\rcoeffp{\square},\outcoeffp{}=\outcoeffp{\square}$
to obtain the cross sections. However, this is not suitable for numerical
evaluation with truncation in multipole degree; hence we need to express
them in terms of particle-wise expansions $\rcoeffp p,\outcoeffp p$.
The original incident field re-expanded around $p$-th particle reads
according to (\ref{eq:reqular vswf coefficient vector translation})
\begin{equation}
\rcoeffincp p=\troprp p{\square}\rcoeffp{\square}\label{eq:a_inc local from global}
\end{equation}
whereas the contributions of fields scattered from each particle expanded
around the global origin $\vect r_{\square}$ is, according to (\ref{eq:singular to regular vswf coefficient vector translation}),
\begin{equation}
\outcoeffp{\square}=\sum_{q\in\mathcal{P}}\troprp{\square}q\outcoeffp q.\label{eq:f global from local}
\end{equation}
Using the unitarity (\ref{eq:regular translation unitarity}) and
composition (\ref{eq:regular translation composition}) properties,
one has
\begin{align}
\rcoeffp{\square}^{\dagger}\outcoeffp{\square} & =\rcoeffincp p^{\dagger}\troprp p{\square}\troprp{\square}q\outcoeffp q=\rcoeffincp p^{\dagger}\sum_{q\in\mathcal{P}}\troprp pqf_{q}\nonumber \\
 & =\sum_{q\in\mathcal{P}}\left(\troprp qp\rcoeffincp p\right)^{\dagger}f_{q}=\sum_{q\in\mathcal{P}}\rcoeffincp q^{\dagger}f_{q},\label{eq:atf form multiparticle}
\end{align}
where only the last expression is suitable for numerical evaluation
with truncated matrices, because the previous ones contain a translation
operator right next to an incident field coefficient vector. Similarly,
\begin{align}
\left\Vert \outcoeffp{\square}\right\Vert ^{2} & =\outcoeffp{\square}^{\dagger}\outcoeffp{\square}=\sum_{p\in\mathcal{P}}\left(\troprp{\square}p\outcoeffp p\right)^{\dagger}\sum_{q\in\mathcal{P}}\troprp{\square}q\outcoeffp q\nonumber \\
 & =\sum_{p\in\mathcal{P}}\sum_{q\in\mathcal{P}}\outcoeffp p^{\dagger}\troprp pq\outcoeffp q.\label{eq:f squared form multiparticle}
\end{align}
Substituting (\ref{eq:atf form multiparticle}), (\ref{eq:f squared form multiparticle})
into (\ref{eq:scattering CS single}) and (\ref{eq:absorption CS single}),
we get the many-particle expressions for extinction, scattering and
absorption cross sections suitable for numerical evaluation: 
\begin{eqnarray}
\sigma_{\mathrm{ext}}\left(\uvec k,\hat{\vect E}_{0}\right) & = & -\frac{1}{\kappa^{2}\left\Vert \vect E_{0}\right\Vert ^{2}}\Re\sum_{p\in\mathcal{P}}\rcoeffincp p^{\dagger}\outcoeffp p=-\frac{1}{2k^{2}\left\Vert \vect E_{0}\right\Vert ^{2}}\Re\sum_{p\in\mathcal{P}}\rcoeffincp p^{\dagger}\left(\Tp p+\Tp p^{\dagger}\right)\rcoeffp p,\label{eq:extincion CS multi}\\
\sigma_{\mathrm{scat}}\left(\uvec k,\hat{\vect E}_{0}\right) & = & \frac{1}{\kappa^{2}\left\Vert \vect E_{0}\right\Vert ^{2}}\sum_{p\in\mathcal{P}}\sum_{q\in\mathcal{P}}\outcoeffp p^{\dagger}\troprp pq\outcoeffp q\nonumber \\
 &  & =\frac{1}{\kappa^{2}\left\Vert \vect E_{0}\right\Vert ^{2}}\sum_{p\in\mathcal{P}}\sum_{q\in\mathcal{P}}\rcoeffp p^{\dagger}\Tp p^{\dagger}\troprp pq\Tp q\rcoeffp q,\label{eq:scattering CS multi}\\
\sigma_{\mathrm{abs}}\left(\uvec k,\hat{\vect E}_{0}\right) & = & -\frac{1}{\kappa^{2}\left\Vert \vect E_{0}\right\Vert ^{2}}\sum_{p\in\mathcal{P}}\Re\left(\outcoeffp p^{\dagger}\left(\rcoeffincp p+\sum_{q\in\mathcal{P}}\troprp pq\outcoeffp q\right)\right).\nonumber \\
\label{eq:absorption CS multi}
\end{eqnarray}
An alternative approach to derive the absorption cross section is
via a power transport argument. Note the direct proportionality between
absorption cross section (\ref{eq:absorption CS single}) and net
radiated power for single scatterer (\ref{eq:Power transport}), $\sigma_{\mathrm{abs}}=-\eta_{0}\eta P/2\left\Vert \vect E_{0}\right\Vert ^{2}$.
In the many-particle setup (with non-lossy background medium, so that
only the particles absorb), the total absorbed power is equal to the
sum of absorbed powers on each particle, $-P=\sum_{p\in\mathcal{P}}-P_{p}$.
Using the power transport formula (\ref{eq:Power transport}) particle-wise
gives
\begin{equation}
\sigma_{\mathrm{abs}}\left(\uvec k\right)=-\frac{1}{\kappa^{2}\left\Vert \vect E_{0}\right\Vert ^{2}}\sum_{p\in\mathcal{P}}\left(\Re\left(\rcoeffp p^{\dagger}\outcoeffp p\right)+\left\Vert \outcoeffp p\right\Vert ^{2}\right)\label{eq:absorption CS multi alternative}
\end{equation}
which seems different from (\ref{eq:absorption CS multi}), but using
(\ref{eq:particle total incident field coefficient a}), we can rewrite
it as
\begin{align*}
\sigma_{\mathrm{abs}}\left(\uvec k\right) & =-\frac{1}{\kappa^{2}\left\Vert \vect E_{0}\right\Vert ^{2}}\sum_{p\in\mathcal{P}}\Re\left(\outcoeffp p^{\dagger}\left(\rcoeffp p+\outcoeffp p\right)\right)\\
 & =-\frac{1}{\kappa^{2}\left\Vert \vect E_{0}\right\Vert ^{2}}\sum_{p\in\mathcal{P}}\Re\left(\outcoeffp p^{\dagger}\left(\rcoeffincp p+\sum_{q\in\mathcal{P}\backslash\left\{ p\right\} }\tropsp pq\outcoeffp q+\outcoeffp p\right)\right).
\end{align*}
It is easy to show that all the terms of $\sum_{p\in\mathcal{P}}\sum_{q\in\mathcal{P}\backslash\left\{ p\right\} }\outcoeffp p^{\dagger}\tropsp pq\outcoeffp q$
containing the singular spherical Bessel functions $y_{l}$ are imaginary,
 so that actually $\sum_{p\in\mathcal{P}}\Re\left(\sum_{q\in\mathcal{P}\backslash\left\{ p\right\} }\outcoeffp p^{\dagger}\tropsp pq\outcoeffp q+\outcoeffp p^{\dagger}\outcoeffp p\right)=\sum_{p\in\mathcal{P}}\Re\left(\sum_{q\in\mathcal{P}\backslash\left\{ p\right\} }\outcoeffp p^{\dagger}\troprp pq\outcoeffp q+\outcoeffp p^{\dagger}\outcoeffp p\right)=\sum_{p\in\mathcal{P}}\Re\left(\sum_{q\in\mathcal{P}\backslash\left\{ p\right\} }\outcoeffp p^{\dagger}\troprp pq\outcoeffp q+\outcoeffp p^{\dagger}\troprp pp\outcoeffp p\right)=\sum_{p\in\mathcal{P}}\sum_{q\in\mathcal{P}}\outcoeffp p^{\dagger}\troprp pq\outcoeffp q,$
proving that the expressions in (\ref{eq:absorption CS multi}) and
(\ref{eq:absorption CS multi alternative}) are equal.

\section{Infinite periodic systems
\global\long\def\dlv{\protect\vect a}%
\global\long\def\rlv{\protect\vect b}%
}\label{sec:Infinite}

Although large finite systems are where MSTMM excels the most, there
are several reasons that makes its extension to infinite lattices
(where periodic boundary conditions might be applied) desirable as
well. Other methods might be already fast enough, but MSTMM will be
faster in most cases in which there is enough spacing between the
neighboring particles. MSTMM works well with any space group symmetry
the system might have (as opposed to, for example, FDTD with a cubic
mesh applied to a honeycomb lattice), which makes e.g. application
of group theory in mode analysis quite easy. And finally, having
a method that handles well both infinite and large finite systems
gives a possibility to study finite-size effects in periodic scatterer
arrays.

\subsection{Formulation of the problem}\label{subsec:Quasiperiodic scattering problem}

Let us have a linear system of compact EM scatterers on a homogeneous
background as in Section (\ref{subsec:Multiple-scattering}), but
this time, the system shall be periodic: let there be a $d$-dimensional
($d$ can be 1, 2 or 3) Bravais lattice embedded into the three-dimensional
real space, with lattice vectors $\left\{ \dlv_{i}\right\} _{i=1}^{d}$,
and let the lattice points be labeled with an $d$-dimensional integer
multi-index $\vect n\in\ints^{d}$, so the lattice points have cartesian
coordinates $\vect R_{\vect n}=\sum_{i=1}^{d}n_{i}\vect a_{i}$. There
can be several scatterers per unit cell with indices $\alpha$ from
a set $\mathcal{P}_{1}$ and (relative) positions $\vect r_{\alpha}$
inside the unit cell; any particle of the periodic system can thus
be labeled by a multi-index from $\mathcal{P}=\ints^{d}\times\mathcal{P}_{1}$.
The scatterers are located at positions $\vect r_{\vect n,\alpha}=\vect R_{\vect n}+\vect r_{\alpha}$
and their $T$-matrices are periodic, $T_{\vect n,\alpha}=T_{\alpha}$.
In such system, the multiple-scattering problem (\ref{eq:Multiple-scattering problem})
can be rewritten as

\begin{equation}
\outcoeffp{\vect n,\alpha}-T_{\alpha}\sum_{\left(\vect m,\beta\right)\in\mathcal{P}\backslash\left\{ \left(\vect n,\alpha\right)\right\} }\tropsp{\vect n,\alpha}{\vect m,\beta}\outcoeffp{\vect m,\beta}=T_{\alpha}\rcoeffincp{\vect n,\alpha}.\quad\left(\vect n,\alpha\right)\in\mathcal{P}\label{eq:Multiple-scattering problem periodic}
\end{equation}

Due to periodicity, we can also write $\tropsp{\vect n,\alpha}{\vect m,\beta}=\tropsp{\alpha}{\beta}\left(\vect R_{\vect m}-\vect R_{\vect n}\right)=\tropsp{\alpha}{\beta}\left(\vect R_{\vect m-\vect n}\right)=\tropsp{\vect 0,\alpha}{\vect m-\vect n,\beta}$.
Assuming quasi-periodic right-hand side with quasi-momentum $\vect k$,
$\rcoeffincp{\vect n,\alpha}=\rcoeffincp{\vect 0,\alpha}\left(\vect k\right)e^{i\vect k\cdot\vect R_{\vect n}}$,
the solutions of (\ref{eq:Multiple-scattering problem periodic})
will be also quasi-periodic according to Bloch theorem, $\outcoeffp{\vect n,\alpha}=\outcoeffp{\vect 0,\alpha}\left(\vect k\right)e^{i\vect k\cdot\vect R_{\vect n}}$,
and eq.\ (\ref{eq:Multiple-scattering problem periodic}) can be
rewritten as follows
\begin{align}
\outcoeffp{\vect 0,\alpha}\left(\vect k\right)e^{i\vect k\cdot\vect R_{\vect n}}-T_{\alpha}\sum_{\left(\vect m,\beta\right)\in\mathcal{P}\backslash\left\{ \left(\vect n,\alpha\right)\right\} }\tropsp{\vect n,\alpha}{\vect m,\beta}\outcoeffp{\vect 0,\beta}\left(\vect k\right)e^{i\vect k\cdot\vect R_{\vect m}} & =T_{\alpha}\rcoeffincp{\vect 0,\alpha}\left(\vect k\right)e^{i\vect k\cdot\vect R_{\vect n}},\nonumber \\
\outcoeffp{\vect 0,\alpha}\left(\vect k\right)-T_{\alpha}\sum_{\left(\vect m,\beta\right)\in\mathcal{P}\backslash\left\{ \left(\vect n,\alpha\right)\right\} }\tropsp{\vect 0,\alpha}{\vect m-\vect n,\beta}\outcoeffp{\vect 0,\beta}\left(\vect k\right)e^{i\vect k\cdot\vect R_{\vect m-\vect n}} & =T_{\alpha}\rcoeffincp{\vect 0,\alpha}\left(\vect k\right),\nonumber \\
\outcoeffp{\vect 0,\alpha}\left(\vect k\right)-T_{\alpha}\sum_{\left(\vect m,\beta\right)\in\mathcal{P}\backslash\left\{ \left(\vect 0,\alpha\right)\right\} }\tropsp{\vect 0,\alpha}{\vect m,\beta}\outcoeffp{\vect 0,\beta}\left(\vect k\right)e^{i\vect k\cdot\vect R_{\vect m}} & =T_{\alpha}\rcoeffincp{\vect 0,\alpha}\left(\vect k\right),\nonumber \\
\outcoeffp{\vect 0,\alpha}\left(\vect k\right)-T_{\alpha}\sum_{\beta\in\mathcal{P}}W_{\alpha\beta}\left(\vect k\right)\outcoeffp{\vect 0,\beta}\left(\vect k\right) & =T_{\alpha}\rcoeffincp{\vect 0,\alpha}\left(\vect k\right),\label{eq:Multiple-scattering problem unit cell}
\end{align}
so we reduced the initial scattering problem to one involving only
the field expansion coefficients from a single unit cell, but we need
to compute the ``lattice Fourier transform'' of the translation
operator,
\begin{equation}
W_{\alpha\beta}(\vect k)\equiv\sum_{\vect m\in\ints^{d}}\left(1-\delta_{\alpha\beta}\delta_{\vect m\vect 0}\right)\tropsp{\vect 0,\alpha}{\vect m,\beta}e^{i\vect k\cdot\vect R_{\vect m}},\label{eq:W definition}
\end{equation}
evaluation of which is possible but rather non-trivial due to the
infinite lattice sum, so we cover it separately in Sect. (\ref{subsec:W operator evaluation}).

As in the case of a finite system, eq.\ (\ref{eq:Multiple-scattering problem unit cell})
can be written in a shorter block-matrix form,
\begin{equation}
\left(I-TW\right)\outcoeffp{\vect 0}\left(\vect k\right)=T\rcoeffincp{\vect 0}\left(\vect k\right)\label{eq:Multiple-scattering problem unit cell block form}
\end{equation}
 Eq.\ (\ref{eq:Multiple-scattering problem unit cell}) can be used
to calculate electromagnetic response of the structure to external
quasiperiodic driving field -- most notably a plane wave. However,
the non-trivial solutions of the equation with right hand side (i.e.
the external driving) set to zero, 
\begin{equation}
\left(I-TW\right)\outcoeffp{\vect 0}\left(\vect k\right)=0,\label{eq:lattice mode equation}
\end{equation}
describes the \emph{lattice modes}, i.e. electromagnetic excitations
that can sustain themselves for prolonged time even without external
driving\emph{.} Non-trivial solutions to (\ref{eq:lattice mode equation})
exist if the matrix on the left-hand side $M\left(\omega,\vect k\right)=\left(I-T\left(\omega\right)W\left(\omega,\vect k\right)\right)$
is singular -- this condition gives the \emph{dispersion relation}
for the periodic structure. Note that in realistic (lossy) systems,
at least one of the pair $\omega,\vect k$ will acquire complex values.
The solution $\outcoeffp{\vect 0}\left(\vect k\right)$ is then obtained
as the right  singular vector of $M\left(\omega,\vect k\right)$
corresponding to the zero singular value.

Loss in the scatterers causes the solutions of (\ref{eq:lattice mode equation})
shift to complex frequencies. If the background medium has constant
real refractive index $n$, negative (or positive) imaginary part
of the frequency $\omega$ causes an artificial gain (or loss) in
the medium, which manifests itself as exponential magnification (or
attenuation) of the radial parts of the translation operators, $h_{l}^{\left(1\right)}\left(rn\omega/c\right)$,
w.r.t.\ the distance; the gain might then balance the losses in particles,
resulting in sustained modes satisfying eq.\ (\ref{eq:lattice mode equation}).

\subsection{Numerical solution}

In practice, equation (\ref{eq:Multiple-scattering problem unit cell block form})
is solved in the same way as eq.\ (\ref{eq:Multiple-scattering problem block form})
in the multipole degree truncated form. The lattice mode problem (\ref{eq:lattice mode equation})
is (after multipole degree truncation) solved by finding $\omega,\vect k$
for which the matrix $M\left(\omega,\vect k\right)$ has a zero singular
value. A naïve approach to do that is to sample a volume with a grid
in the $\left(\omega,\vect k\right)$ space, performing a singular
value decomposition of $M\left(\omega,\vect k\right)$ at each point
and finding where the lowest singular value of $M\left(\omega,\vect k\right)$
is close enough to zero. However, this approach is quite expensive,
since $W\left(\omega,\vect k\right)$ has to be evaluated for each
$\omega,\vect k$ pair separately (unlike the original finite case
(\ref{eq:Multiple-scattering problem block form}) translation operator
$\trops$, which, for a given geometry, depends only on frequency).
Therefore, a much more efficient but not completely robust approach
to determine the photonic bands is to sample the $\vect k$-space
(a whole Brillouin zone or its part) and for each fixed $\vect k$
to find a corresponding frequency $\omega$ with zero singular value
of $M\left(\omega,\vect k\right)$ using a minimisation algorithm
(two- or one-dimensional, depending on whether one needs the exact
complex-valued $\omega$ or whether the its real-valued approximation
is satisfactory). Typically, a good initial guess for $\omega\left(\vect k\right)$
is obtained from the empty lattice approximation, $\left|\vect k\right|=\sqrt{\epsilon\mu}\omega/c_{0}$
(modulo reciprocal lattice points). A somehow challenging step is
to distinguish the different bands that can all be very close to the
empty lattice approximation, especially if the particles in the system
are small. In high-symmetry points of the Brilloin zone, this can
be solved by factorising $M\left(\omega,\vect k\right)$ into irreducible
representations $\Gamma_{i}$ and performing the minimisation in each
irrep separately, cf. Section \ref{sec:Symmetries}, and using the
different $\omega_{\Gamma_{i}}\left(\vect k\right)$ to obtain the
initial guesses for the nearby points $\vect k+\delta\vect k$. 

\begin{figure}
\begin{centering}
\includegraphics[width=1\columnwidth]{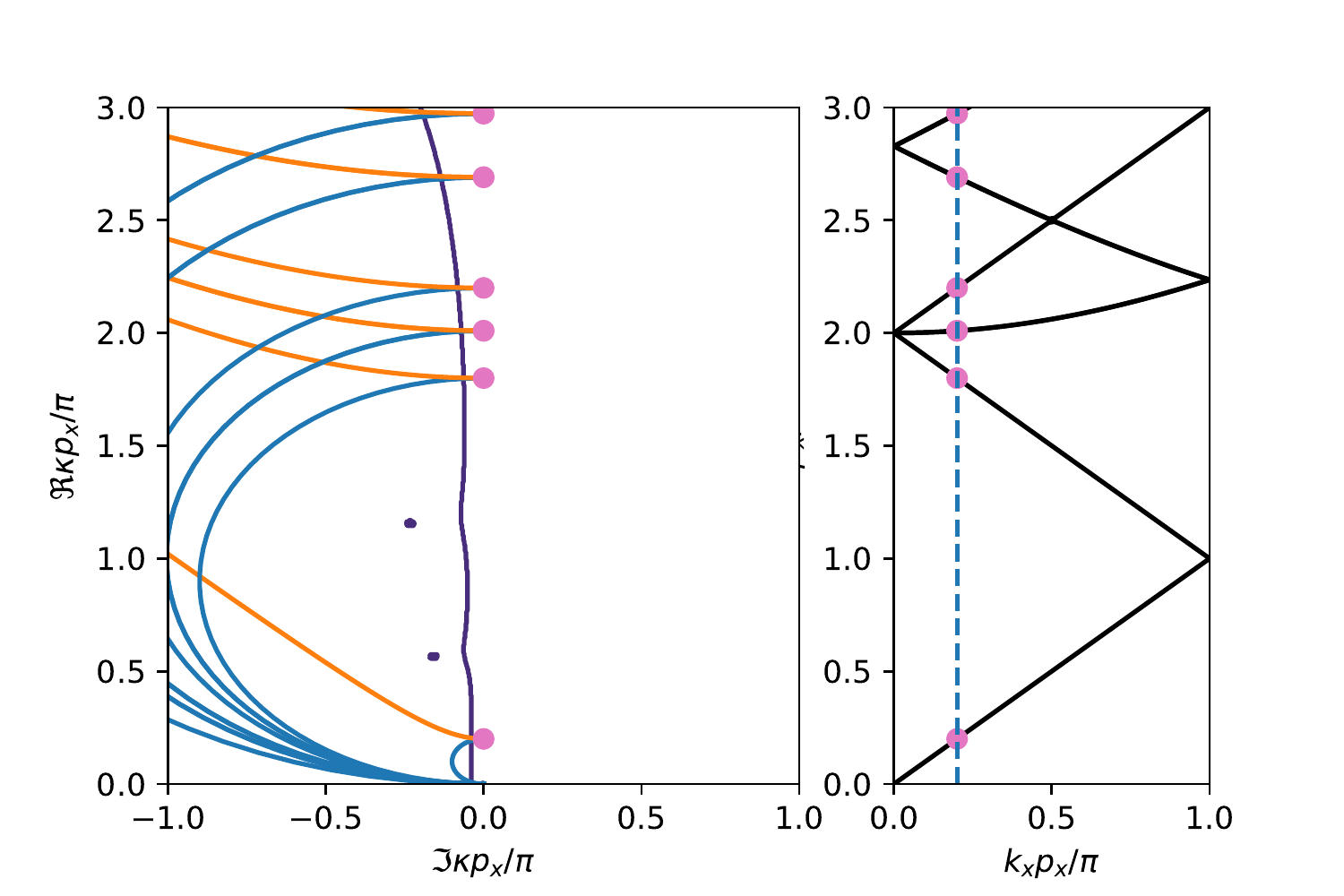}
\par\end{centering}
\caption{Left: Illustration of branch cuts in $M\left(\omega,\protect\vect k\right)$
obtained using Ewald summation over two-dimensional square lattice
in three-dimensional space filled with dielectric medium with constant
real refraction index $n$ and wavenumber $\kappa\left(\omega\right)=\omega n/c$.
The function is holomorphic in the positive imaginary half-plane.
The points corresponding to the diffraction orders of an ``empty''
lattice lie on the real axis (pink), and from each of them two branch
cuts originate: one due to the branch cut in the incomplete $\Gamma$
function (orange, hyperbolic shape), and another due to the branch
cut of $\gamma(z)$ if the branch is selected to be continuous for
$-3\pi/2<\arg\left(z-1\right)<\pi/2$  (blue, circular shape). Further
non-analyticities might stem from the material model: the violet curve
represents a branch cut originating from a complex square root in
the refractive index $n_{\mathrm{Au}}\left(\omega\right)=\sqrt{\varepsilon_{\mathrm{Au}}\left(\omega\right)}$,
where $\varepsilon_{\mathrm{Au}}\left(\omega\right)$ is the Drude-Lorentz
permittivity model of gold used for the scatterers. The other parameters
used here are $p_{x}=580\,\mathrm{nm}$ (lattice period), $\protect\vect k=\left(0.2\pi/p_{x},0\right)$,
$n=1.52$. The plot on the right shows the ``empty'' lattice diffraction
orders on the line $\protect\vect k=\left(k_{x},0\right),k_{x}\in\left[0,\pi/p_{x}\right].$
}\label{fig:ewald branch cuts}
\end{figure}

An alternative, faster and more robust approach to generic minimisation
algorithms are eigensolvers for nonlinear eigenvalue problems based
on contour integration \cite{beyn_integral_2012,gavin_feast_2018}
which are able to find the roots of $M\left(\omega,\vect k\right)=0$
inside an area enclosed by a given complex frequency plane contour,
assuming that $M\left(\omega,\vect k\right)$ is an analytical function
of $\omega$ inside the contour. A necessary prerequisite for this
is that all the ingredients of $M\left(\omega,\vect k\right)$ are
analytical as well. It practice, this usually means that interpolation
cannot be used in a straightforward way for material properties or
$T$-matrices. For material response, constant permittivity or Drude-Lorentz
models suit this purpose well. The need to evaluate the $T$-matrices
precisely (without the speedup provided by interpolation) at many
points might cause a performance bottleneck for scatterers with more
complicated shapes. And finally, the integration contour has to evade
any branch cuts appearing in the lattice-summed translation operator
$W\left(\omega,\vect k\right)$, as described in the following and
illustrated in Fig.\ \ref{fig:ewald branch cuts}.

\subsection{Computing the lattice sum of the translation operator}\label{subsec:W operator evaluation}

The problem in evaluating (\ref{eq:W definition}) is the asymptotic
behaviour of the translation operator at large distances, $\tropsp{\vect 0,\alpha}{\vect m,\beta}\sim\left|\vect R_{\vect m}\right|^{-1}e^{i\kappa\left|\vect R_{\vect m}\right|}$,
so that its lattice sum does not in the strict sense converge for
any $d>1$ -dimensional lattice unless $\Im\kappa>0$. The problem
of poorly converging lattice sums can be solved by decomposing the
lattice-summed function into two parts: a short-range part that decays
fast and can be summed directly, and a long-range part which decays
poorly but is fairly smooth everywhere, so that its Fourier transform
decays fast enough, and to deal with the long range part by Poisson
summation over the reciprocal lattice; these two parts put together
shall give an analytical continuation of the original sum for $\Im\kappa\le0$.
This idea dates back to Ewald \cite{ewald_berechnung_1921} who solved
the problem for electrostatic potentials (Green's functions for Laplace's
equation). For linear electrodynamic problems, ruled by Helmholtz
equation, the same basic idea can be used as well, resulting in exponentially
convergent summation formulae, but the technical details are considerably
more complicated than in electrostatics. For the scalar Helmholtz
equation in three dimensions, the formulae for lattice Green's functions
were developed by Ham \& Segall \cite{ham_energy_1961} for 3D periodicity,
Kambe \cite{kambe_theory_1967,kambe_theory_1967-1,kambe_theory_1968}
for 2D periodicity and Moroz \cite{moroz_quasi-periodic_2006} for
1D periodicity. A review of these methods can be found in \cite{linton_lattice_2010}. 

For our purposes we do not need directly the lattice Green's functions
but rather the related lattice sums of spherical wavefunctions defined
in (\ref{eq:sigma lattice sums}), which can be derived by an analogous
procedure. Below, we state the results in a form independent upon
the normalisation and phase conventions for spherical harmonic bases
(pointing out some errors in the aforementioned literature) and discuss
some practical aspects of the numerical evaluation. The derivation
of the somewhat more complicated 1D and 2D periodicities is provided
in the Supplementary Material. 

We note that the lattice sums for \emph{scalar} Helmholtz equation
are enough for the evaluation of the translation operator lattice
sum $W_{\alpha\beta}(\vect k)$: in eq.\ (\ref{eq:translation operator singular})
we demonstratively expressed the translation operator elements as
linear combinations of (outgoing) \emph{scalar} spherical wavefunctions
\begin{equation}
\sswfoutlm lm\left(\vect r\right)=h_{l}^{\left(1\right)}\left(r\right)\ush lm\left(\uvec r\right).\label{eq:scalar spherical wavefunctions}
\end{equation}
If we formally label 
\begin{equation}
\sigma_{l,m}\left(\vect k,\vect s\right)=\sum_{\vect n\in\ints^{d}}\left(1-\delta_{\vect{R_{n}},\vect s}\right)e^{i\vect{\vect k}\cdot\vect R_{\vect n}}\sswfoutlm lm\left(\kappa\left(\vect s+\vect{R_{n}}\right)\right),\label{eq:sigma lattice sums}
\end{equation}
we see from eqs. (\ref{eq:translation operator singular}),(\ref{eq:W definition})
that the matrix elements of $W_{\alpha\beta}(\vect k)$ read  
\[
W_{\alpha,\tau lm;\beta,\tau'l'm'}(\vect k)=\sum_{\lambda=\left|l-l'\right|+\left|\tau-\tau'\right|}^{l+l'}\tropcoeff_{\tau lm;\tau'l'm'}^{\lambda}\sigma_{\lambda,m-m'}\left(-\vect k,\vect r_{\alpha}-\vect r_{\beta}\right),\quad\tau'\ne\tau,
\]
where the constant factors are exactly the same as in (\ref{eq:translation operator constant factors}).

The lattice sums $\sigma_{l,m}\left(\vect k,\vect s\right)$ are related
to what is also called \emph{structural constants} in some literature
\cite{kambe_theory_1967,kambe_theory_1967-1,kambe_theory_1968}, but
the phase and normalisation differ. For reader's reference, we list
the Ewald-type formulae for lattice sums $\sigma_{l,m}\left(\vect k,\vect s\right)$
rewritten in a way that is independent on particular phase or normalisation
conventions of vector spherical harmonics. 

In all three lattice dimensionality cases, the lattice sums are divided
into short-range and long-range parts, $\sigma_{l,m}\left(\vect k,\vect s\right)=\sigma_{l,m}^{\left(\mathrm{S},\eta\right)}\left(\vect k,\vect s\right)+\sigma_{l,m}^{\left(\mathrm{L},\eta\right)}\left(\vect k,\vect s\right)$
depending on a positive parameter $\eta$. The short-range part has
in all three cases the same form:

\begin{multline}
\sigma_{l,m}^{\left(\mathrm{S},\eta\right)}\left(\vect k,\vect s\right)=-\frac{2^{l+1}i}{\kappa^{l+1}\sqrt{\pi}}\sum_{\vect n\in\ints^{d}}\left(1-\delta_{\vect{R_{n}},-\vect s}\right)\left|\vect s_{\vect n}\right|^{l}\ush lm\left(\uvec s_{\vect n}\right)e^{i\vect k\cdot\vect{R_{n}}}\\
\times\int_{\eta}^{\infty}e^{-\left|\vect s_{\vect n}\right|^{2}\xi^{2}}e^{-\kappa^{2}/4\xi^{2}}\xi^{2l}\ud\xi\\
+\delta_{\vect{R_{n}},-\vect s}\frac{\delta_{l0}\delta_{m0}}{\sqrt{4\pi}}\Gamma\left(-\frac{1}{2},-\frac{\kappa^{2}}{4\eta^{2}}\right)\ush lm\left(\uvec s_{\vect n}\right),\label{eq:Ewald in 3D short-range part}
\end{multline}
where we labeled $\vect s_{\vect n}\equiv\vect s+\vect R_{\vect n}$.
The formal $\left(1-\delta_{\vect{R_{n}},-\vect s}\right)$ factor
here accounts for leaving out the direct excitation of a particle
by itself, corresponding to the $\left(1-\delta_{\alpha\beta}\delta_{\vect m\vect 0}\right)$
factor in (\ref{eq:W definition}). The leaving out then causes an
additional (``self-interaction'') term on the last line of (\ref{eq:Ewald in 3D short-range part}),
which appears only when the displacement vector $\vect s$ coincides
with a lattice point. Strictly speaking, this is not a ``short-range''
term, hence it is often noted separately in the literature; however,
we keep it in $\sigma_{l,m}^{\left(\mathrm{S},\eta\right)}\left(\vect k,\vect s\right)$
for formal convenience. $\Gamma(a,z)$ is the incomplete Gamma function.

In practice, the integrals in (\ref{eq:Ewald in 3D short-range part})
can be easily evaluated by numerical quadrature and the incomplete
$\Gamma$-functions using the series or continued fraction representations
from \cite{NIST:DLMF}.

The explicit form of the long-range part of the lattice sum depends
on the lattice dimensionality. The long-range parts are calculated
as sums over the reciprocal lattice $\Lambda^{*}$ with lattice vectors
$\left\{ \vect b_{i}\right\} _{i=1}^{d}$ lying in the same $d$-dimensional
subspace as the direct lattice vectors $\left\{ \vect a_{i}\right\} _{i=1}^{d}$
and satisfying $\vect a_{i}\cdot\vect b_{j}=\delta_{ij}$. In the
following, let us label $\vect k_{\vect K}\equiv\vect k+\vect K$,
where $\vect K$ is a point in the reciprocal lattice, and let $\mathcal{A}$
be the lattice unit cell volume (or area/length in the 2D/1D cases).

\paragraph{Case $d=3$}

\begin{equation}
\sigma_{l,m}^{\left(\mathrm{L},\eta\right)}\left(\vect k,\vect s\right)=\frac{4\pi i^{l+1}}{\kappa\mathcal{A}}\sum_{\vect K\in\Lambda^{*}}e^{-i\vect k_{\vect K}\cdot\vect s}\frac{\left(\left|\vect k_{\vect K}\right|/\kappa\right)^{l}}{\kappa^{2}-\left|\vect k_{\vect K}\right|^{2}}e^{\left(\kappa^{2}-\left|\vect k_{\vect K}\right|^{2}\right)/4\eta^{2}}\ush lm\left(\uvec k_{\vect K}\right)\label{eq:Ewald in 3D long-range part 3D}
\end{equation}
regardless of chosen coordinate axes. Here $\mathcal{A}$ is the unit
cell volume (or length/area in the following 1D/2D lattice cases).

\paragraph{Cases $d=1,2$}

In the quasiperiodic cases, we decompose vectors into parallel and
orthogonal parts with respect to the linear subspace in which the
Bravais lattice lies (the reciprocal lattice lies in the same subspace),
$\vect v=\vect v_{\perp}+\vect v_{\parallel}$, and we label 
\begin{equation}
\gamma_{\vect k_{\vect K}}\equiv\gamma_{\vect k_{\vect K}}\left(\kappa\right)\equiv\left(\left|\vect k_{\vect K}\right|^{2}-\kappa^{2}\right)^{\frac{1}{2}}/\kappa,\label{eq:lilgamma}
\end{equation}
\begin{equation}
\Delta_{d;j}\left(x,z\right)\equiv\int_{x}^{\infty}t^{-\frac{d_{c}}{2}-n}\exp\left(-t+\frac{z^{2}}{4t}\right)\ud t,\label{eq:Delta_j}
\end{equation}
where $d_{c}=3-d$ is the complementary dimension of the lattice.
Then
\begin{multline}
\sigma_{l}^{m}\left(\vect k,\vect s\right)=\frac{-i}{2\pi^{d_{c}/2}\mathcal{A}\kappa}\frac{\left(2l+1\right)!!}{\kappa^{l}}\sum_{\vect K\in\Lambda^{*}}e^{-i\vect k_{\vect K}\cdot\vect s}\times\\
\times\sum_{j=0}^{l}\frac{\left(-1\right)^{j}}{j!}\left(\frac{\kappa\gamma_{\vect k_{\vect K}}}{2}\right)^{2j}\Delta_{d;j}\left(\frac{\kappa^{2}\gamma_{\vect k_{\vect K}}^{2}}{4\eta^{2}},-i\kappa\gamma_{\vect k_{\vect K}}\left|\vect s_{\perp}\right|\right)\times\\
\times\sum_{l'=\max\left(0,l-2j\right)}^{l-j}4\pi i^{l'}\left(2\left|\vect s_{\bot}\right|\right)^{2j-l+l'}\frac{\left|\vect k_{\vect K}\right|^{l'}}{\left(2l'+1\right)!!}\sum_{m'=-l'}^{l'}\ush{l'}{m'}\left(\uvec k_{\vect K}\right)\times\\
\times\int\ud\Omega_{\vect r}\,\ush lm\left(\uvec r\right)\ushD{l'}{m'}\left(\uvec r\right)\left(\frac{\left|\vect r_{\perp}\right|}{\left|\vect r\right|}\right)^{l-l}\left(\frac{-\vect r_{\perp}\cdot\vect s_{\perp}}{\left|\vect r_{\perp}\right|\left|\vect s_{\perp}\right|}\right)^{2j-l+l'}.\label{eq:Ewald in 3D long-range part 1D 2D}
\end{multline}
The angular integral on the last line of (\ref{eq:Ewald in 3D long-range part 1D 2D})
gives a set of constant coefficients characteristic to a chosen convention
for spherical harmonics and coordinate axes; relatively simple closed-form
expressions are obtained for 2D periodicity if we choose the lattice
to lie in the $xy$ plane, so that both $\vect r_{\perp},\vect s_{\perp}$
are parallel to the $z$ axis, as done in \cite{kambe_theory_1968},
see also Supplementary Material. In the special case $\vect s_{\perp}=0$
the expressions can be considerably simplified as most of the terms
vanish and $\Delta_{d;j}\left(x,0\right)=\Gamma\left(1-d_{c}/2-j,x\right)$,
but the general case is needed for evaluating the fields in space
(see Section \ref{subsec:Periodic scattering and fields}) or if there
is an offset between two particles in a unitcell that is not parallel
to the lattice subspace.

If $s_{\bot}\ne0$, the integral $\Delta_{d;j}\left(x,0\right)$ can
be evaluated e.g.\ using the Taylor series\foreignlanguage{finnish}{
\[
\Delta_{d;j}\left(x,z\right)=\sum_{k=0}^{\infty}\Gamma\left(1-\frac{d_{c}}{2}-j-k,x\right)\frac{\left(z/2\right)^{2k}}{k!}
\]
which has infinite radius of convergence and is the first choice for
small $z$}. Kambe \cite{kambe_theory_1968} mentions a recurrence
formula that can be obtained integrating (\ref{eq:Delta_j}) by parts
(with signs corrected here):
\begin{equation}
\Delta_{d;j+1}\left(x,z\right)=\frac{4}{z^{2}}\left(\left(\frac{1}{2}-j\right)\Delta_{d;j}\left(x,z\right)-\Delta_{d;j-1}\left(x,z\right)+x^{\frac{d_{c}}{2}-j}e^{-x+\frac{z^{2}}{4x}}\right)\label{eq:Delta_j recurrent}
\end{equation}
with the first two terms for 2D periodicity
\begin{align*}
\Delta_{2;0}\left(x,z\right) & =\frac{\sqrt{\pi}}{2}e^{-x^{2}+\frac{z^{2}}{4x}}\left(w\left(-\frac{z}{2\sqrt{x}}+i\sqrt{x}\right)+w\left(\frac{z}{2\sqrt{x}}+i\sqrt{x}\right)\right),\\
\Delta_{2;1}\left(x,z\right) & =i\frac{\sqrt{\pi}}{z}e^{-x^{2}+\frac{z^{2}}{4x}}\left(w\left(-\frac{z}{2\sqrt{x}}+i\sqrt{x}\right)-w\left(\frac{z}{2\sqrt{x}}+i\sqrt{x}\right)\right),
\end{align*}
where $w\left(z\right)=e^{-z^{2}}\left(1+2i\pi^{-1/2}\int_{0}^{z}e^{t^{2}}\ud t\right)$
is the Faddeeva function. However, the recurrence formula (\ref{eq:Delta_j recurrent})
is unsuitable for numerical evaluation if $z$ is small or $j$ is
large due to its numerical instability.

One pecularity of the two-dimensional case is the two-branchedness
of $\gamma_{\vect k_{\vect K}}\left(\kappa\right)$ and the incomplete
$\Gamma$-function $\Gamma\left(\frac{1}{2}-j,z\right)$ appearing
in the long-range part (in the cases $d=1,3$ the function $\gamma_{\vect k_{\vect K}}\left(\kappa\right)$
appears with even powers, and $\Gamma\left(-j,z\right)$ is meromorphic
for integer $j$  \cite[8.2.9]{NIST:DLMF}). As a consequence, if
we now explicitly label the dependence on the wavenumber, $\sigma_{l,m}^{\left(\mathrm{L},\eta\right)}\left(\kappa,\vect k,\vect s\right)$
has branch points at $\kappa=\left|\vect k+\vect K\right|$ for every
reciprocal lattice vector $\vect K$. If the wavenumber $\kappa$
of the medium has a positive imaginary part, $\Im\kappa>0$, then
the translation operator elements $\trops_{\tau lm;\tau'l'm}\left(\kappa\vect r\right)$
decay exponentially as $\left|\vect r\right|\to\infty$ and the lattice
sum in (\ref{eq:W definition}) converges absolutely even in the direct
space, and it is equal to the Ewald sum with the principal branches
used both in $\gamma\left(z\right)$ and $\Gamma\left(\frac{1}{2}-j,z\right)$
\cite{linton_lattice_2010}. For other values of $\kappa$, we typically
choose the branch in such way that $W_{\alpha\beta}\left(\vect k\right)$
is analytically continued even when the wavenumber's imaginary part
crosses the real axis. The principal value of $\Gamma\left(\frac{1}{2}-j,z\right)$
has a branch cut at the negative real half-axis, which, considering
the lattice sum as a function of $\kappa$, translates into branch
cuts starting at $\kappa=\left|\vect k+\vect K\right|$ and continuing
in straight lines towards $+\infty$. Therefore, in the quadrant $\Re z<0,\Im z\ge0$
we use the continuation of the principal value from $\Re z<0,\Im z<0$
instead of the principal branch \cite[8.2.9]{NIST:DLMF}, moving the
branch cut in the $z$ variable to the positive imaginary half-axis.
This moves the branch cuts w.r.t.\ $\kappa$ away from the real axis,
as illustrated in Fig.\ \ref{fig:ewald branch cuts}.  

\subsubsection{Choice of Ewald parameter and high-frequency breakdown}

The Ewald parameter $\eta$ determines the pace of convergence of
both parts. The larger $\eta$ is, the faster $\sigma_{l,m}^{\left(\mathrm{S},\eta\right)}\left(\vect k,\vect s\right)$
converges but the slower $\sigma_{l,m}^{\left(L,\eta\right)}\left(\vect k,\vect s\right)$
converges. Therefore (based on the lattice geometry) it has to be
adjusted in a way that a reasonable amount of terms needs to be evaluated
numerically from both $\sigma_{l,m}^{\left(\mathrm{S},\eta\right)}\left(\vect k,\vect s\right)$
and $\sigma_{l,m}^{\left(\mathrm{L},\eta\right)}\left(\vect k,\vect s\right)$.
For one-dimensional, square, and cubic lattices, the optimal choice
for small frequencies (wavenumbers) is $\eta=\sqrt{\pi}/p$ where
$p$ is the direct lattice period \cite{linton_lattice_2010}. However,
in floating point arithmetics, the magnitude of the summands must
be taken into account as well in order to maintain accuracy.

There is a particular problem with the ``central'' reciprocal lattice
points in the long-range sums for which the real part of $\left|\vect k_{\vect K}\right|^{2}-\kappa^{2}$
is negative: the incomplete $\Gamma$ function present in the sum
(either explicitly or in the expansions of $\Delta_{j}$) grows exponentially
with respect to the negative second argument, with asymptotic behaviour
$\Gamma\left(a,z\right)\sim e^{-z}z^{a-1}$ . Therefore for higher
frequencies, the parameter $\eta$ needs to be adjusted in a way that
keeps the value of $\Gamma\left(a,\left(\left|\vect k_{\vect K}\right|^{2}-\kappa^{2}\right)/4\eta^{2}\right)$
within reasonable bounds. If we assume that $\vect k$ lies in the
first Brillouin zone, the minimum real part of the second argument
of the $\Gamma$ function will be $\left(\left|\vect k\right|^{2}-\kappa^{2}\right)/4\eta^{2}$,
so setting $\eta\ge\sqrt{\left|\kappa\right|^{2}-\left|\vect k\right|^{2}}/2\log M$
eliminates the exponential growth in the incomplete $\Gamma$ function,
where the constant $M$ is chosen to represent the (rough) maximum
tolerated magnitude of the summand with regard to target accuracy.
This adjustment means that, in the worst-case scenario, with growing
wavenumber one has to include an increasing number of terms in the
long-range sum in order to achieve a given accuracy, the number of
terms being proportional to $\left|\kappa\right|^{d}$ where $d$
is the dimension of the lattice.

\subsection{Scattering cross sections and field intensities in periodic system}\label{subsec:Periodic scattering and fields}

Once the scattering (\ref{eq:Multiple-scattering problem unit cell block form})
or mode problem (\ref{eq:lattice mode equation}) is solved, one can
evaluate some useful related quantities, such as scattering cross
sections (coefficients) or field intensities.

For plane wave scattering on 2D lattices, one can directly use the
formulae (\ref{eq:extincion CS multi}), (\ref{eq:absorption CS multi alternative}),
taking the sums over scatterers inside one unit cell, to get the extinction
and absorption cross sections per unit cell. From these, quantities
such as absorption, extinction and scattering coefficients are obtained
using suitable normalisation by unit cell size, depending on lattice
dimensionality.

Ewald summation can be used for evaluating scattered field intensities
outside scatterers' circumscribing spheres: this requires expressing
VSWF cartesian components in terms of scalar spherical wavefunctions
defined in (\ref{eq:scalar spherical wavefunctions}). Fortunately,
these can be obtained easily from the expressions for the translation
operator: 
\begin{align}
\vswfrtlm{\tau}lm\left(\kappa\vect r\right) & =\sum_{m'=-1}^{1}\tropr_{\tau lm;21m'}\left(\kappa\vect r\right)\vswfrtlm 21{m'}\left(0\right),\nonumber \\
\vswfouttlm{\tau}lm\left(\kappa\vect r\right) & =\sum_{m'=-1}^{1}\trops_{\tau lm;21m'}\left(\kappa\vect r\right)\vswfrtlm 21{m'}\left(0\right),\label{eq:VSWFs expressed as translated dipole waves}
\end{align}
which follows from eqs. (\ref{eq:regular vswf translation}), (\ref{eq:singular vswf translation})
and the fact that all the other regular VSWFs except for $\vswfrtlm 21{m'}$
vanish at origin. For the quasiperiodic scattering problem formulated
in section \ref{subsec:Quasiperiodic scattering problem}, the total
electric field scattered from all the particles at point $\vect r$
located outside all the particles' circumscribing sphere reads, using
eqs. (\ref{eq:translation operator singular}), (\ref{eq:sigma lattice sums}),
(\ref{eq:scalar spherical wavefunctions}),
\begin{multline}
\vect E_{\mathrm{scat}}\left(\vect r\right)=\sum_{\left(\vect n,\alpha\right)\in\mathcal{P}}\sum_{\tau lm}\outcoeffptlm{\vect n,\alpha}{\tau}lm\vect u_{\tau lm}\left(\kappa\left(\vect r-\text{\ensuremath{\vect R_{\vect n}}-\ensuremath{\vect r_{\alpha}}}\right)\right)=\\
=\sum_{\alpha\in\mathcal{P}_{1}}\sum_{\tau lm}\outcoeffptlm{\vect 0,\alpha}{\tau}lm\sum_{m'=-1}^{1}\vswfrtlm 21{m'}\left(0\right)\sum_{\lambda=\left|l-1\right|+\left|\tau-2\right|}^{l+1}\tropcoeff_{\tau lm;21m'}^{\lambda}\sigma_{\lambda,m-m'}\left(-\vect k,\vect r-\vect r_{\alpha}\right).\label{eq:Scattered fields in periodic systems}
\end{multline}
In the scattering problem, the total field intensity is obtained by
adding the incident field to (\ref{eq:Scattered fields in periodic systems});
whereas in the lattice mode problem the total field is directly given
by (\ref{eq:Scattered fields in periodic systems}).

\section{Symmetries}\label{sec:Symmetries}

If the system has nontrivial point group symmetries, group theory
gives additional understanding of the system properties, and can be
used to substantially reduce the computational costs.

As an example, if the system has a $D_{2h}$ symmetry and the corresponding
truncated $\left(I-T\trops\right)$ matrix has size $N\times N$,
it can be block-diagonalized into eight blocks of size about $N/8\times N/8$,
each of which can be LU-factorised separately (this is due to the
fact that $D_{2h}$ has eight different one-dimensional irreducible
representations). This can reduce both memory and time requirements
to solve the scattering problem (\ref{eq:Multiple-scattering problem block form})
by a factor of 64.

In periodic systems (problems (\ref{eq:Multiple-scattering problem unit cell block form}),
(\ref{eq:lattice mode equation})) due to small number of particles
per unit cell, the costliest part is usually the evaluation of the
lattice sums in the $W\left(\omega,\vect k\right)$ matrix, not the
linear algebra. However, decomposition of the lattice mode problem
(\ref{eq:lattice mode equation}) into the irreducible representations
of the corresponding little co-groups of the system's space group
is nevertheless a useful tool in the mode analysis: among other things,
it enables separation of the lattice modes (which can then be searched
for each irrep separately), and the irrep dimension gives a priori
information about mode degeneracy.

\subsection{Excitation coefficients under point group operations}

In order to make use of the point group symmetries, we first need
to know how they affect our basis functions, i.e.\ the VSWFs. Let
$g$ be a member of the orthogonal group $\mathrm{O}(3)$, i.e.\ a
3D point rotation or reflection operation that transforms vectors
in $\reals^{3}$ with an orthogonal matrix $R_{g}$:
\[
\vect r\mapsto R_{g}\vect r.
\]
With $\groupop g$ we shall denote the action of $g$ on a field in
real space. For a scalar field $w$ we have $\left(\groupop gw\right)\left(\vect r\right)=w\left(R_{g}^{-1}\vect r\right)$,
whereas for a vector field $\vect w$, $\left(\groupop g\vect w\right)\left(\vect r\right)=R_{g}\vect w\left(R_{g}^{-1}\vect r\right)$.

Spherical harmonics $\ush lm$, being a basis of the $l$-dimensional
representation of $\mathrm{O}(3)$, transform as \cite[Chapter 15]{wigner_group_1959}
\begin{equation}
\left(\groupop g\ush lm\right)\left(\uvec r\right)=\ush lm\left(R_{g}^{-1}\uvec r\right)=\sum_{m'=-l}^{l}D_{m,m'}^{l}\left(g\right)\ush l{m'}\left(\uvec r\right)\label{eq:Wigner matrices}
\end{equation}
where $D_{m,m'}^{l}\left(g\right)$ denotes the elements of the \emph{Wigner
matrix } representing the operation $g$. From their definitions
(\ref{eq:vector spherical harmonics definition}) and the properties
of the gradient operator under coordinate transforms, vector spherical
harmonics $\vsh 2lm,\vsh 3lm$ transform in the same way,
\begin{align*}
\left(\groupop g\vsh 2lm\right)\left(\uvec r\right) & =\sum_{m'=-l}^{l}D_{m,m'}^{l}\left(g\right)\vsh 2l{m'}\left(\uvec r\right),\\
\left(\groupop g\vsh 3lm\right)\left(\uvec r\right) & =\sum_{m'=-l}^{l}D_{m,m'}^{l}\left(g\right)\vsh 3l{m'}\left(\uvec r\right),
\end{align*}
but the remaining set $\vsh 1lm$ transforms differently due to their
pseudovector nature stemming from the cross product in their definition:
\[
\left(\groupop g\vsh 1lm\right)\left(\uvec r\right)=\sum_{m'=-l}^{l}\widetilde{D_{m,m'}^{l}}\left(g\right)\vsh 1l{m'}\left(\uvec r\right),
\]
where $\widetilde{D_{m,m'}^{l}}\left(g\right)=D_{m,m'}^{l}\left(g\right)$
if $g$ is a proper rotation, $g\in\mathrm{SO(3)}$, but for spatial
inversion operation $i:\vect r\mapsto-\vect r$ we have $D_{m,m'}^{l}\left(i\right)=\left(-1\right)^{l}$
but $\widetilde{D_{m,m'}^{l}}\left(i\right)=\left(-1\right)^{l+1}$.
The transformation behaviour of vector spherical harmonics directly
propagates to vector spherical waves, cf. (\ref{eq:VSWF regular}),
(\ref{eq:VSWF outgoing}):
\begin{align*}
\left(\groupop g\vswfouttlm 1lm\right)\left(\vect r\right) & =\sum_{m'=-l}^{l}\widetilde{D_{m,m'}^{l}}\left(g\right)\vswfouttlm 1l{m'}\left(\vect r\right),\\
\left(\groupop g\vswfouttlm 2lm\right)\left(\vect r\right) & =\sum_{m'=-l}^{l}D_{m,m'}^{l}\left(g\right)\vswfouttlm 2l{m'}\left(\vect r\right),
\end{align*}
and analogously for the regular waves $\vswfrtlm{\tau}lm$.  For
convenience, we introduce the symbol $D_{m,m'}^{\tau l}$ that describes
the transformation of both (``magnetic'' and ``electric'') types
of waves at once:
\[
\groupop g\vswfouttlm{\tau}lm\left(\vect r\right)=\sum_{m'=-l}^{l}D_{m,m'}^{\tau l}\left(g\right)\vswfouttlm{\tau}l{m'}\left(\vect r\right).
\]
Note that this symbol retains the unitarity of the original Wigner
matrices,
\begin{equation}
\sum_{m'}\left(D_{m,m'}^{\tau l}\left(g\right)\right)^{*}D_{\mu,m'}^{\tau l}\left(g\right)=\delta_{m\mu}.\label{eq:Wigner matrix unitarity}
\end{equation}
Using these, we can express the VSWF expansion (\ref{eq:E field expansion})
of the electric field around origin in a rotated/reflected system,
\begin{equation}
\left(\groupop g\vect E\right)\left(\omega,\vect r\right)=\sum_{\tau=1,2}\sum_{l=1}^{\infty}\sum_{m=-l}^{+l}\sum_{m'=-l}^{l}\left(\rcoefftlm{\tau}lmD_{m,m'}^{\tau l}\left(g\right)\vswfrtlm{\tau}l{m'}\left(\kappa\vect r\right)+\outcoefftlm{\tau}lmD_{m,m'}^{\tau l}\left(g\right)\vswfouttlm{\tau}l{m'}\left(\kappa\vect r\right)\right),\label{eq:Group action on electric field VSWF expansion}
\end{equation}
which, together with the $T$-matrix definition, (\ref{eq:T-matrix definition})
can be used to obtain a $T$-matrix of a rotated or mirror-reflected
particle. Let $T$ be the $T$-matrix of an original particle; the
$T$-matrix of a particle physically transformed by operation $g\in O(3)$
is then (following from eqs. (\ref{eq:Group action on electric field VSWF expansion}),
(\ref{eq:T-matrix definition}), (\ref{eq:Wigner matrix unitarity}))
\begin{equation}
T'_{\tau lm;\tau'l'm'}=\sum_{\mu=-l}^{l}\sum_{\mu'=-l'}^{l'}D_{\mu,m}^{\tau l}\left(g\right)T_{\tau l\mu;\tau'l'\mu'}\left(D_{\mu',m'}^{\tau'l'}\left(g\right)\right)^{*}.\label{eq:T-matrix of a transformed particle}
\end{equation}
If the particle is symmetric (so that $g$ produces a particle indistinguishable
from the original one), the $T$-matrix must remain invariant under
the transformation (\ref{eq:T-matrix of a transformed particle}),
$T'_{\tau lm;\tau'l'm'}=T{}_{\tau lm;\tau'l'm'}$. Explicit forms
of these invariance properties for the most imporant point group symmetries
can be found in \cite{schulz_point-group_1999}.

If the field expansion is done around a point $\vect r_{p}$ different
from the global origin, as in \ref{eq:E field expansion multiparticle},
we have
\begin{multline}
\left(\groupop g\vect E\right)\left(\omega,\vect r\right)=\sum_{\tau=1,2}\sum_{l=1}^{\infty}\sum_{m=-l}^{+l}\sum_{m'=-l}^{l}\left(\rcoeffptlm p{\tau}lmD_{m,m'}^{\tau l}\left(g\right)\vswfrtlm{\tau}l{m'}\left(\kappa\left(\vect r-R_{g}\vect r_{p}\right)\right)\right.+\\
+\left.\outcoeffptlm p{\tau}lmD_{m,m'}^{\tau l}\left(g\right)\vswfouttlm{\tau}l{m'}\left(\kappa\left(\vect r-R_{g}\vect r_{p}\right)\right)\right).\label{eq:rotated E field expansion around outside origin}
\end{multline}

\begin{figure}
\begin{centering}
\tikzstyle{orbit1}=[rectangle,draw=blue!50,fill=blue!20]
\tikzstyle{orbit2}=[rectangle,draw=green!50,fill=green!20]
\begin{tikzpicture}
	\draw (-4,0) -- (4,0);
	\draw (0,-2) -- (0,2);

	\node (O) at (0,0) [circle, draw=red!50,fill=red!20] {O};

		\node (A) at (-3.0,0.5) [orbit1] {A};
		\node (B) at (3.0,0.5) [orbit1] {B};
		\node (C) at (3.0,-0.5) [orbit1] {C};
		\node (D) at (-3.0,-0.5) [orbit1] {D};

		\node (E) at (0,-1) [orbit2] {E};
		\node (F) at (0,1)  [orbit2] {F};

	\draw [->] (A) to (B);
	\draw [->] (A) to [bend right=18] (C);
	\draw [->] (A) to (D);

	\draw [<->] (E.east) to [bend right=30] (F.east);

\end{tikzpicture}
\par\end{centering}
\caption{Scatterer orbits under $D_{2}$ symmetry. Particles $A,B,C,D$ lie
outside of origin or any mirror planes, and together constitute an
orbit of the size equal to the order of the group, $\left|D_{2}\right|=4$.
Particles $E,F$ lie on the $yz$ plane, hence the corresponding reflection
maps each of them to itself, but the $xz$ reflection (or the $\pi$
rotation around the $z$ axis) maps them to each other, forming a
particle orbit of size 2. The particle $O$ in the very origin is
always mapped to itself, constituting its own orbit.}\label{fig:D2-symmetric structure particle orbits}
\end{figure}
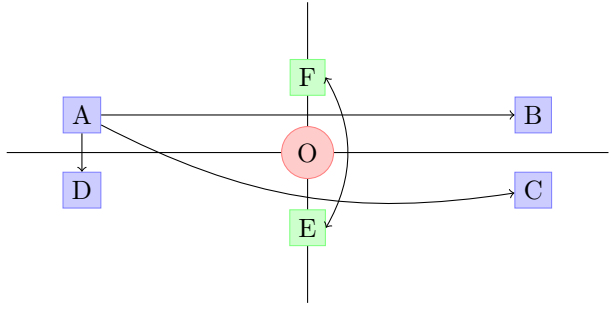

With these transformation properties in hand, we can proceed to the
effects of point symmetries on the whole many-particle system. Let
us have a many-particle system symmetric with respect to a point group
$G$. A symmetry operation $g\in G$ determines a permutation of the
particles: $p\mapsto\pi_{g}(p)$, $p\in\mathcal{P}$; their positions
transform as $\vect r_{\pi_{g}p}=R_{g}\vect r_{p}$, $\vect r_{\pi_{g}^{-1}p}=R_{g}^{-1}\vect r_{p}$.
In the symmetric multiple-scattering problem, transforming the whole
field according to $g$, in terms of field expansion around a particle
originally labelled as $p$
\begin{align*}
\left(\groupop g\vect E\right)\left(\omega,\vect r\right) & =\sum_{\tau=1,2}\sum_{l=1}^{\infty}\sum_{m=-l}^{+l}\sum_{m'=-l}^{l}\left(\rcoeffptlm p{\tau}lmD_{m,m'}^{\tau l}\left(g\right)\vswfrtlm{\tau}l{m'}\left(\kappa\left(\vect r-R_{g}\vect r_{p}\right)\right)\right.+\\
 & \quad+\left.\outcoeffptlm p{\tau}lmD_{m,m'}^{\tau l}\left(g\right)\vswfouttlm{\tau}l{m'}\left(\kappa\left(\vect r-R_{g}\vect r_{p}\right)\right)\right)\\
 & =\sum_{\tau=1,2}\sum_{l=1}^{\infty}\sum_{m=-l}^{+l}\sum_{m'=-l}^{l}\left(\rcoeffptlm p{\tau}lmD_{m,m'}^{\tau l}\left(g\right)\vswfrtlm{\tau}l{m'}\left(\kappa\left(\vect r-\vect r_{\pi_{g}p}\right)\right)\right.\\
 & \quad+\left.\outcoeffptlm p{\tau}lmD_{m,m'}^{\tau l}\left(g\right)\vswfouttlm{\tau}l{m'}\left(\kappa\left(\vect r-\vect r_{\pi_{g}p}\right)\right)\right)\\
 & =\sum_{\tau=1,2}\sum_{l=1}^{\infty}\sum_{m=-l}^{+l}\sum_{m'=-l}^{l}\left(\rcoeffptlm{\pi_{g}^{-1}q}{\tau}lmD_{m,m'}^{\tau l}\left(g\right)\vswfrtlm{\tau}l{m'}\left(\kappa\left(\vect r-\vect r_{q}\right)\right)\right.\\
 & \quad+\left.\outcoeffptlm{\pi_{g}^{-1}q}{\tau}lmD_{m,m'}^{\tau l}\left(g\right)\vswfouttlm{\tau}l{m'}\left(\kappa\left(\vect r-\vect r_{q}\right)\right)\right).
\end{align*}
In the last step, we relabeled $q=\pi_{g}p$. This means that the
field expansion coefficients $\rcoeffp p,\outcoeffp p$ transform
as 
\begin{align}
\rcoeffptlm p{\tau}l{m'} & \overset{g}{\longmapsto}\sum_{m=-l}^{l}\rcoeffptlm{\pi_{g}^{-1}(p)}{\tau}lmD_{m,m'}^{\tau l}\left(g\right),\nonumber \\
\outcoeffptlm p{\tau}l{m'} & \overset{g}{\longmapsto}\sum_{m=-l}^{l}\outcoeffptlm{\pi_{g}^{-1}(p)}{\tau}lmD_{m,m'}^{\tau l}\left(g\right).\label{eq:excitation coefficient under symmetry operation}
\end{align}
For a given particle $p$, we will call the set of particles onto
which any of the symmetries maps the particle $p$,\ i.e. the set
$\left\{ \pi_{g}\left(p\right);g\in G\right\} $, as the \emph{orbit}
of particle $p$. The whole set $\mathcal{P}$ can therefore be divided
into the different particle orbits; an example is in Fig. \ref{fig:D2-symmetric structure particle orbits}.
The importance of the particle orbits stems from fact that the expansion
coefficients belonging to particles in different orbits are not related
together under the group action in (\ref{eq:excitation coefficient under symmetry operation}).
As before, we introduce a short-hand pairwise matrix notation for
(\ref{eq:excitation coefficient under symmetry operation}) 
\begin{align}
\rcoeffp p & \overset{g}{\longmapsto}\tilde{J}\left(g\right)\rcoeffp{\pi_{g}^{-1}(p)},\nonumber \\
\outcoeffp p & \overset{g}{\longmapsto}\tilde{J}\left(g\right)\outcoeffp{\pi_{g}^{-1}(p)},\label{eq:excitation coefficient under symmetry operation matrix form}
\end{align}
and also a global block-matrix form

\begin{align}
\rcoeff & \overset{g}{\longmapsto}J\left(g\right)a,\nonumber \\
\outcoeff & \overset{g}{\longmapsto}J\left(g\right)\outcoeff.\label{eq:excitation coefficient under symmetry operation global block form}
\end{align}
If the particle indices are ordered in a way that the particles belonging
to the same orbit are grouped together, $J\left(g\right)$ will be
a block-diagonal unitary matrix, each block (also unitary) representing
the action of $g$ on one particle orbit. All the $J\left(g\right)$s
make together a (reducible) linear representation of $G$.

\subsection{Irrep decomposition}

Knowledge of symmetry group actions $J\left(g\right)$ on the field
expansion coefficients give us the possibility to construct a symmetry
adapted basis in which we can block-diagonalise the multiple-scattering
problem matrix $\left(I-TS\right)$. Let $\Gamma_{n}$ be the $d_{n}$-dimensional
irreducible matrix representations of $G$ consisting of matrices
$D^{\Gamma_{n}}\left(g\right)$. Then the projection operators
\[
P_{kl}^{\left(\Gamma_{n}\right)}\equiv\frac{d_{n}}{\left|G\right|}\sum_{g\in G}\left(D^{\Gamma_{n}}\left(g\right)\right)_{kl}^{*}J\left(g\right),\quad k,l=1,\dots,d_{n}
\]
project the full scattering system field expansion coefficient vectors
$\rcoeff,\outcoeff$ onto a subspace corresponding to the irreducible
representation $\Gamma_{n}$. The projectors can be used to construct
a unitary transformation $U$ with components
\begin{equation}
U_{nri;p\tau lm}=\frac{d_{n}}{\left|G\right|}\sum_{g\in G}\left(D^{\Gamma_{n}}\left(g\right)\right)_{rr}^{*}J\left(g\right)_{p'\tau'l'm'(nri);p\tau lm}\label{eq:SAB unitary transformation operator}
\end{equation}
where $r$ goes from $1$ to $d_{n}$ and $i$ goes from 1 to the
multiplicity of irreducible representation $\Gamma_{n}$ in the (reducible)
representation of $G$ spanned by the field expansion coefficients
$\rcoeff$ or $\outcoeff$. The indices $p',\tau',l',m'$ are given
by an arbitrary bijective mapping $\left(n,r,i\right)\mapsto\left(p',\tau',l',m'\right)$
with the constraint that for given $n,r,i$ there are at least some
non-zero elements $U_{nri;p\tau lm}$. For details, we refer the reader
to textbooks about group representation theory, e.g. \cite[Chapter 4]{dresselhaus_group_2008}
or \cite[Chapter 2]{bradley_mathematical_1972}. The transformation
given by $U$ transforms the excitation coefficient vectors $\rcoeff,\outcoeff$
into a new, \emph{symmetry-adapted basis}. 

One can show that if an operator $M$ acting on the excitation coefficient
vectors is invariant under the operations of group $G$, meaning that
\[
\forall g\in G:J\left(g\right)MJ\left(g\right)^{\dagger}=M,
\]
then in the symmetry-adapted basis, $M$ is block diagonal, or more
specifically
\[
M_{\Gamma,r,i;\Gamma',r',j}^{\mathrm{s.a.b.}}=\frac{\delta_{\Gamma\Gamma'}\delta_{ij}}{d_{\Gamma}}\sum_{q}M{}_{\Gamma,r,q;\Gamma',r',q}^{\mathrm{s.a.b.}}.
\]
Both the $T$ and $\trops$ operators (and trivially also the identity
$I$) in (\ref{eq:Multiple-scattering problem block form}) are invariant
under the actions of whole system symmetry group, so $\left(I-T\trops\right)$
is also invariant, hence $U\left(I-T\trops\right)U^{\dagger}$ is
a block-diagonal matrix, and the problem (\ref{eq:Multiple-scattering problem block form})
can be solved for each block separately.

From the computational perspective, it is important to note that $U$
is at least as sparse as $J\left(g\right)$ (which is ``orbit-block''
diagonal), hence the block-diagonalisation can be performed fast.

\subsection{Periodic systems}

Also for periodic systems, $M\left(\omega,\vect k\right)=\left(I-T\left(\omega\right)W\left(\omega,\vect k\right)\right)$
from the left hand side of eqs.\ (\ref{eq:Multiple-scattering problem unit cell block form}),
(\ref{eq:lattice mode equation}) can be block-diagonalised in a similar
manner. Hovewer, in this case, $W\left(\omega,\vect k\right)$ is
in general not invariant under the whole point group symmetry subgroup
of the system geometry due to the $\vect k$ dependence. In other
words, only those point symmetries that the $e^{i\vect k\cdot\vect r}$
modulation does not break are preserved, and no preservation of point
symmetries happens unless $\vect k$ lies somewhere in the high-symmetry
parts of the Brillouin zone. However, the high-symmetry points are
usually the ones of the highest physical interest, for it is where
the band edges are typically located. This subsection does not aim
for an exhaustive treatment of the topic of space groups in physics
(which can be found elsewhere \cite{dresselhaus_group_2008,bradley_mathematical_1972}),
here we rather demonstrate how the group action matrices are generated
on a specific example of a symmorphic space group. 

\begin{figure}
\begin{centering}
\includegraphics[width=0.95\textwidth]{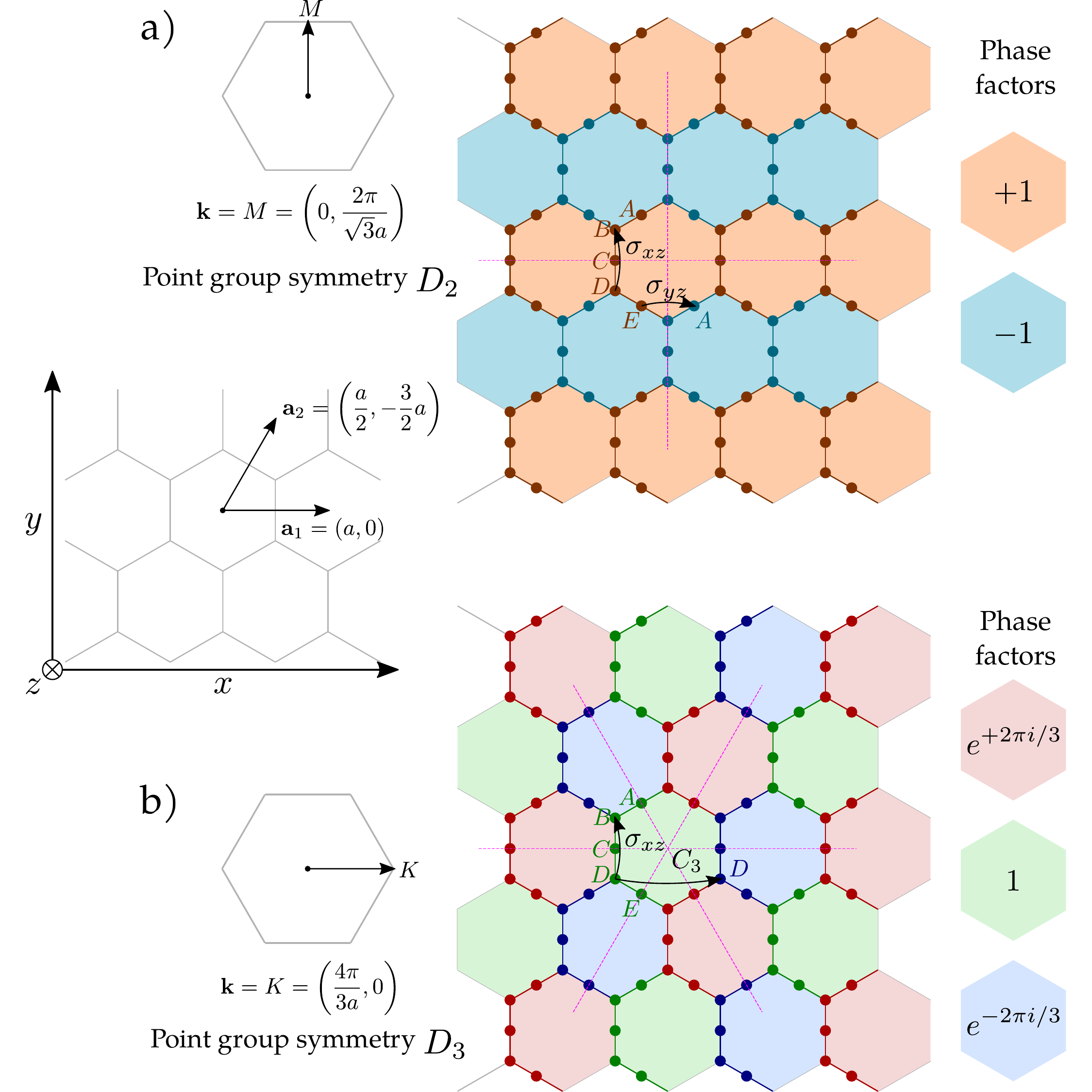}
\par\end{centering}
\caption{Representing symmetry action on electromagnetic Bloch waves in a lattice
with $p6m$ wallpaper group symmetry. In a hexagonal array with five
particles (labeled $A$--$E$) per unit cell, we first choose into
which unit cells do the particles on unit cell boundaries belong.
a) At $M$ point, the little co-group contains a $D_{2}$ point group;
the unit cells can be divided into two groups (alternating horizontal
rows) with opposite sign. The horizontal mirror operation $\sigma_{xz}$
maps the particles from a single unit cell to each other. However,
the vertical mirror operation $\sigma_{yz}$ maps them onto particles
belonging to different unit cells, introducing possible phase factors
in the point group action: $B,C,D$ map onto $B,C,D$ belonging to
different unit cell from same phase group, so no additional phase
is needed; however, $A,E$ map onto $E,A$ belonging to unitcells
with relative phases $\pm\pi$, therefore the corresponding action
matrix blocks will carry a factor $-1$. b) At $K$ point point, little
co-group contains a $D_{3}$ point group, and the unit cells divide
into three groups with relative phase shift $e^{2\pi i/3}$. The horizontal
mirroring $\sigma_{xz}$ again does not introduce any additional phase.
However, the $C_{3}$ rotation mixes particles belonging to different
unit cells, so for example particle $D$ maps onto particle $D$,
but with additional phase factor $e^{-2\pi i/3}$. }\label{Phase factor illustration}
\end{figure}

The transformation to the symmetry adapted basis $U$ is constructed
in a similar way as in the finite case, but because we do not work
with all the (infinite number of) scatterers but only with one unit
cell, additional phase factors $e^{i\vect k\cdot\vect r_{p}}$ appear
in the per-unit-cell group action $J(g)$: this can happen if the
point group symmetry maps some of the scatterers from the reference
unit cell to scatterers belonging to other unit cells. This is illustrated
in Fig.\ \ref{Phase factor illustration}. Fig.\ \ref{Phase factor illustration}a
shows a hexagonal periodic array with $p6m$ wallpaper group symmetry,
with lattice vectors $\vect a_{1}=\left(a,0\right)$ and $\vect a_{2}=\left(a/2,\sqrt{3}a/2\right)$.
We delimit our representative unit cell as the Wigner-Seitz cell with
origin in a $D_{6}$ point group symmetry center (there is one per
each unit cell); per unit cell, there are five different particles
placed on the unit cell boundary, and we need to make a choice to
which unit cell the particles on the boundary belong; in our case,
we choose that a unit cell includes the particles on the left as denoted
by different colors. If the Bloch vector is at the upper $M$ point,
$\vect k=\vect M_{1}=\left(0,2\pi/\sqrt{3}a\right)$, it creates a
relative phase of $\pi$ between the unit cell rows, and the original
$D_{6}$ symmetry is reduced to $D_{2}$. The ``horizontal'' mirror
operation $\sigma_{xz}$ maps, acording to our boundary division,
all the particles only inside the same unit cell, e.g.
\begin{align*}
\outcoeffp{\vect 0A} & \overset{\sigma_{xz}}{\longmapsto}\tilde{J}\left(\sigma_{xz}\right)\outcoeffp{\vect 0E},\\
\outcoeff_{\vect 0C} & \overset{\sigma_{xz}}{\longmapsto}\tilde{J}\left(\sigma_{xz}\right)\outcoeffp{\vect 0C},
\end{align*}
as in eq.\ (\ref{eq:excitation coefficient under symmetry operation}).
However, both the ``vertical'' mirroring $\sigma_{yz}$ and the
$C_{2}$ rotation map the boundary particles onto the boundaries that
do not belong to the reference unit cell with $\vect n=\left(0,0\right)$,
so we have, explicitly writing down also the lattice point indices
$\vect n$,
\begin{align*}
\outcoeffp{\vect 0A} & \overset{\sigma_{yz}}{\longmapsto}\tilde{J}\left(\sigma_{yz}\right)\outcoeffp{\left(0,1\right)E},\\
\outcoeff_{\vect 0C} & \overset{\sigma_{yz}}{\longmapsto}\tilde{J}\left(\sigma_{yz}\right)\outcoeffp{\left(1,0\right)C},
\end{align*}
but we want $J(g)$ to operate only inside one unit cell, so we use
the Bloch condition $\outcoeffp{\vect n,\alpha}=\outcoeffp{\vect 0,\alpha}\left(\vect k\right)e^{i\vect k\cdot\vect R_{\vect n}}$:
in this case, we have $\outcoeffp{\left(0,1\right)\alpha}=\outcoeffp{\vect 0\alpha}e^{i\vect M_{1}\cdot\vect a_{2}}=\outcoeffp{\vect 0\alpha}e^{i0}=\outcoeffp{\vect 0\alpha}$,
$\outcoeffp{\left(1,0\right)\alpha}=e^{i\vect M_{1}\cdot\vect a_{2}}\outcoeffp{\vect 0\alpha}=e^{i\pi}\outcoeffp{\vect 0\alpha}=-\outcoeffp{\vect 0\alpha},$so
\begin{align*}
\outcoeffp{\vect 0A} & \overset{\sigma_{yz}}{\longmapsto}-\tilde{J}\left(\sigma_{yz}\right)\outcoeffp{\vect 0E},\\
\outcoeff_{\vect 0C} & \overset{\sigma_{yz}}{\longmapsto}\tilde{J}\left(\sigma_{yz}\right)\outcoeffp{\vect 0C}.
\end{align*}
If we set instead $\vect k=\vect K=\left(4\pi/3a,0\right),$ the original
$D_{6}$ point group symmetry reduces to $D_{3}$ and the unit cells
can obtain a relative phase factor of $e^{-2\pi i/3}$ (blue) or $e^{2\pi i/3}$
(red). The $\sigma_{xz}$ mirror symmetry, as in the previous case,
acts purely inside the reference unit cell with our boundary division.
However, for a counterclockwise $C_{3}$ rotation as an example we
have (see Fig. \ref{Phase factor illustration}b)
\begin{align*}
\outcoeffp{\vect 0A} & \overset{C_{3}}{\longmapsto}\tilde{J}\left(C_{3}\right)\outcoeffp{\left(0,-1\right)E}=e^{2\pi i/3}\tilde{J}\left(C_{3}\right)\outcoeffp{\vect 0E},\\
\outcoeff_{\vect 0C} & \overset{C_{3}}{\longmapsto}\tilde{J}\left(C_{3}\right)\outcoeffp{\left(1,-1\right)A}=e^{-2\pi i/3}\tilde{J}\left(C_{3}\right)\outcoeffp{\vect 0A},\\
\outcoeff_{\vect 0B} & \overset{C_{3}}{\longmapsto}\tilde{J}\left(C_{3}\right)\outcoeffp{\left(1,-1\right)B}=e^{-2\pi i/3}\tilde{J}\left(C_{3}\right)\outcoeffp{\vect 0B},
\end{align*}
because in this case, the Bloch condition gives $\outcoeffp{\left(0,-1\right)\alpha}=\outcoeffp{\vect 0\alpha}e^{i\vect K\cdot\left(-\vect a_{2}\right)}=\outcoeffp{\vect 0\alpha}e^{-4\pi i/3}=\outcoeffp{\vect 0\alpha}e^{2\pi i/3}=\outcoeffp{\vect 0\alpha}$,
$\outcoeffp{\left(1,-1\right)\alpha}=\outcoeffp{\vect 0\alpha}e^{i\vect K\cdot\left(\vect a_{1}-\vect a_{2}\right)}=e^{-2\pi i/3}\outcoeffp{\vect 0\alpha}.$

Having the group action matrices, we can construct the projectors
and decompose the system into irreducible representations of the corresponding
point groups analogously to the finite case (\ref{eq:SAB unitary transformation operator}).
This procedure can be repeated for any system with a symmorphic space
group symmetry, where the translation and point group operations are
essentially separable. For systems with non-symmorphic space group
symmetries (i.e. those with glide reflection planes or screw rotation
axes) a more refined approach is required \cite{bradley_mathematical_1972,dresselhaus_group_2008}.

\section{Applications}\label{sec:Applications}

Finally, we present some results obtained with the QPMS suite. Scripts
to reproduce these results are available under the \texttt{examples}
directory of the QPMS source repository. 

For further results, used for explaining experiments, see Refs. \cite{hakala_lasing_2017}
(scattering, finite system), \cite{guo_lasing_2019,pourjamal_lasing_2019}
(approximate lattice mode search using a real-frequency-only scan),
and \cite{vakevainen_sub-picosecond_2020} (resonances of a finite
system). Note that in \cite{hakala_lasing_2017,guo_lasing_2019},
$T$-matrices were calculated using a buggy version of SCUFF-EM.

\subsection{Optical response of a square array; finite size effects}

Our first example deals with a plasmonic array made of silver nanoparticles
placed in a square planar configuration. The nanoparticles have shape
of right circular cylinder with 30 nm radius and 30 nm in height.
The particles are placed with periodicity $p_{x}=p_{y}=375\,\mathrm{nm}$
into an isotropic medium with a constant refraction index $n=1.52$.
For silver, we use Drude-Lorentz model with parameters from \cite{rakic_optical_1998},
and the $T$-matrix of a single particle we compute using the null-field
method (with cutoff $l_{\mathrm{max}}=6$ for solving the null-field
equations).  

\begin{figure}
\centering{}\includegraphics[width=0.9\columnwidth]{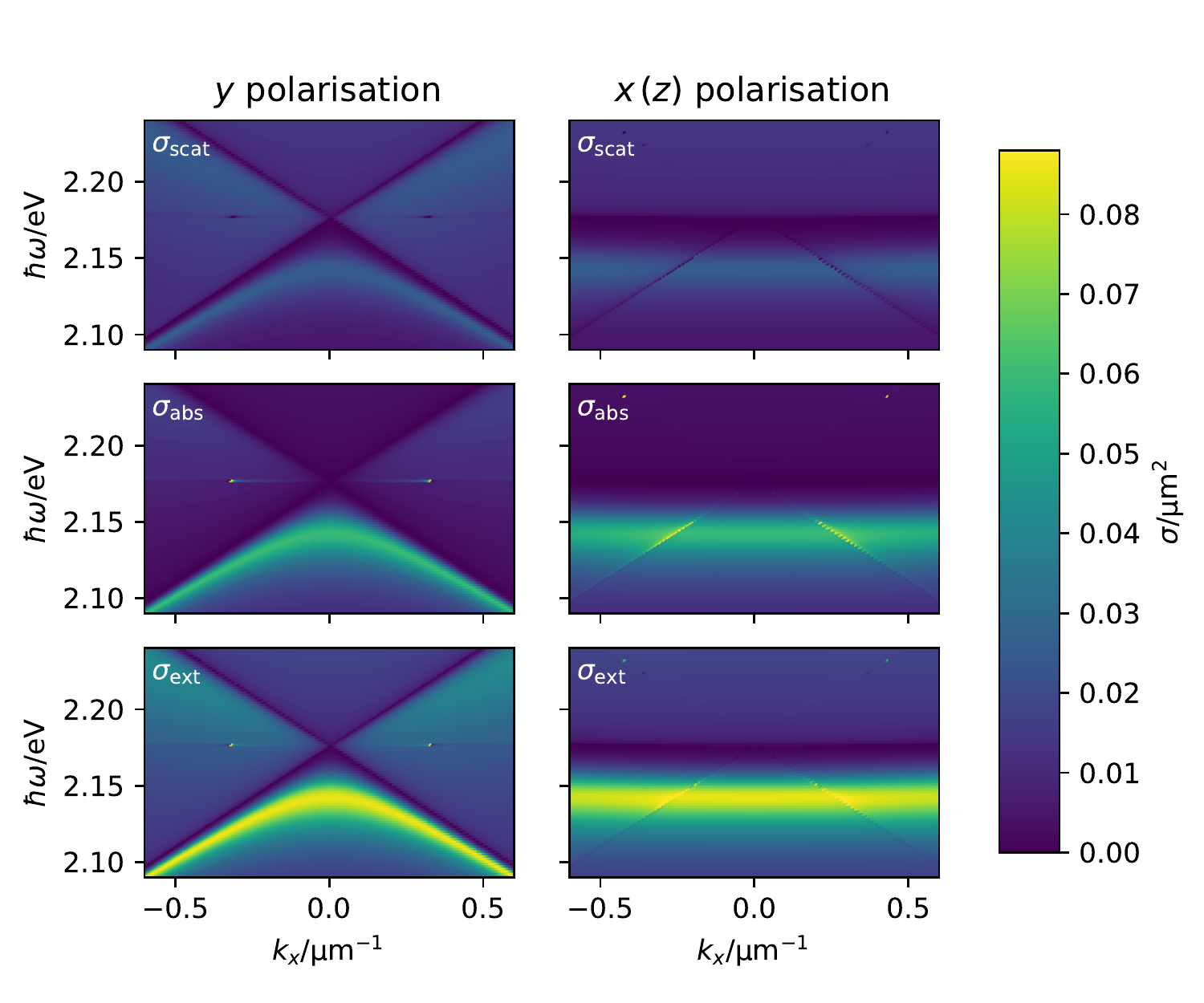}\caption{Response of an infinite square array of silver nanoparticles with
periodicities $p_{x}=p_{y}=375\,\mathrm{nm}$ to plane waves incident
in the $xz$-plane, with $y$-polarised waves (left), and $x$-polarised
waves (right). The images show extinction, scattering and absorption
cross section per unit cell. }\label{fig:Example rectangular absorption infinite}
\end{figure}

\begin{figure}

\begin{centering}
\includegraphics[width=0.9\columnwidth]{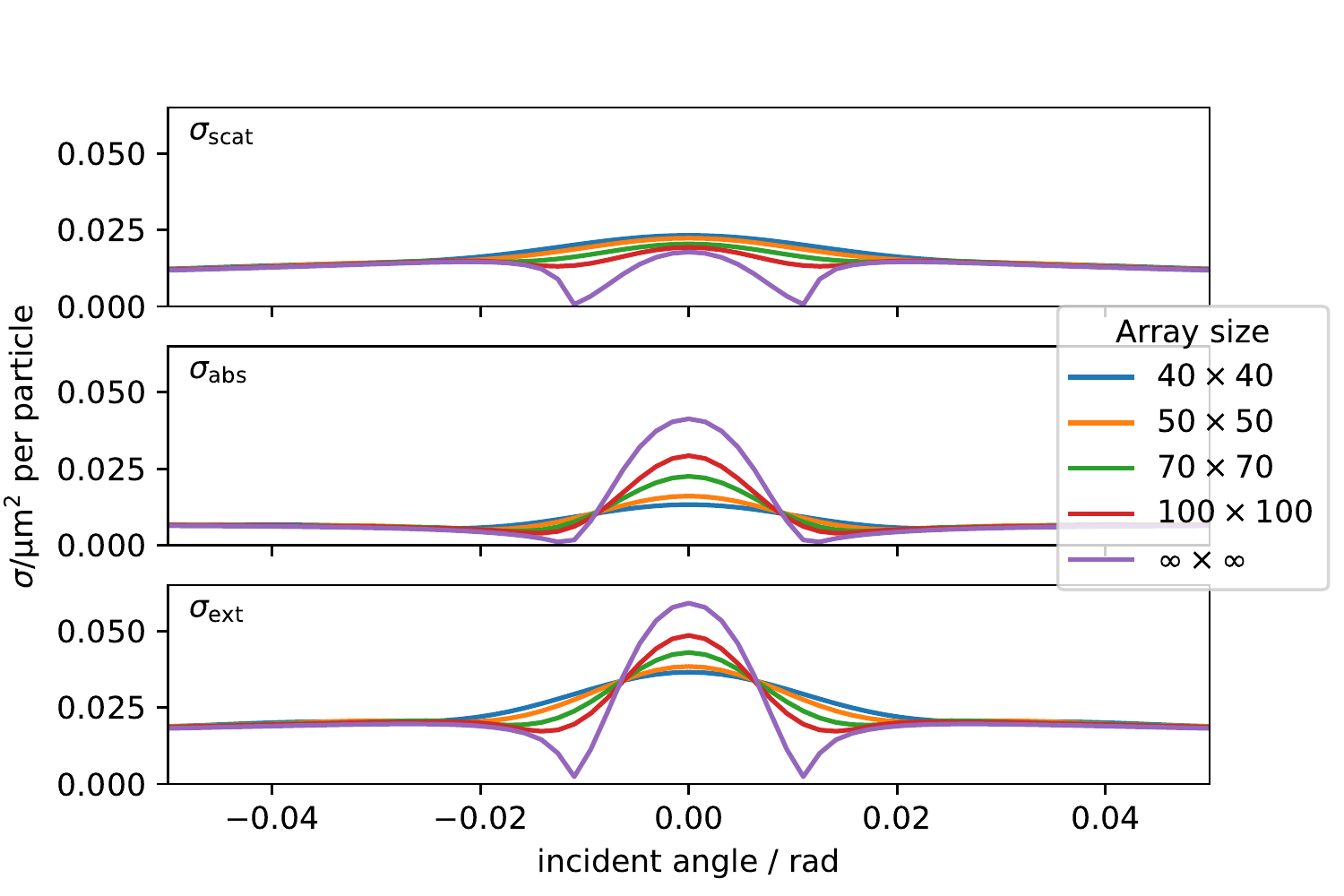}
\par\end{centering}
\caption{Comparison of optical responses of differently sized square arrays
of silver nanoparticles with the same periodicity $p_{x}=p_{y}=375\,\mathrm{nm}$.
In all cases, the array is illuminated by plane waves linearly polarised
in the $y$-direction, with constant frequency $2.15\,\mathrm{eV}/\hbar$.
The cross sections are normalised by the total number of particles
in the array.}\label{fig:Example rectangular absorption size comparison}

\end{figure}

We consider finite arrays with $N_{x}\times N_{y}=40\times40,70\times70,100\times100$
particles and also the corresponding infinite array, and simulate
their absorption when irradiated by  plane waves with incidence direction
lying in the $xz$ plane. We concentrate on the behaviour around the
first diffracted order crossing at the $\Gamma$ point, which happens
around frequency $2.18\,\mathrm{eV}/\hbar$. Figure \ref{fig:Example rectangular absorption infinite}
shows the response for the infinite array for a range of frequencies;
here in particular we used the multipole cutoff $l_{\mathrm{max}}=3$
for the interparticle interactions, although there is no visible difference
if we use $l_{\mathrm{max}}=2$ instead due to the small size of the
particles. In Figure \ref{fig:Example rectangular absorption size comparison},
we compare the response of differently sized array slightly below
the diffracted order crossing. We see that far from the diffracted
orders, all the cross sections are almost directly proportional to
the total number of particles. However, near the resonances, the size
effects become apparent: the lattice resonances tend to fade away
as the size of the array decreases. Moreover, the proportion between
the absorbed and scattered parts changes as while the small arrays
tend to more just scatter the incident light into different directions,
in larger arrays, it is more ``likely'' that the light will scatter
many times, each time sacrifying a part of its energy to the ohmic
losses. 

The finite-size cases in Figure \ref{fig:Example rectangular absorption size comparison}
were computed with quadrupole truncation $l\le2$ and using the decomposition
into the eight irreducible representations of group $D_{2h}$. The
$100\times100$ array took about 4 h to compute on Dell PowerEdge
C4130 with 12 core Xeon E5 2680 v3 2.50GHz, requiring about 20 GB
of RAM. For smaller systems, the computation time decreases quickly,
as the main bottleneck is the LU factorisation. In any case, there
is still room for optimisation in the QPMS suite.

\subsection{Lattice mode structure of a square lattice}

\begin{figure}
\begin{centering}
\includegraphics[width=0.8\columnwidth]{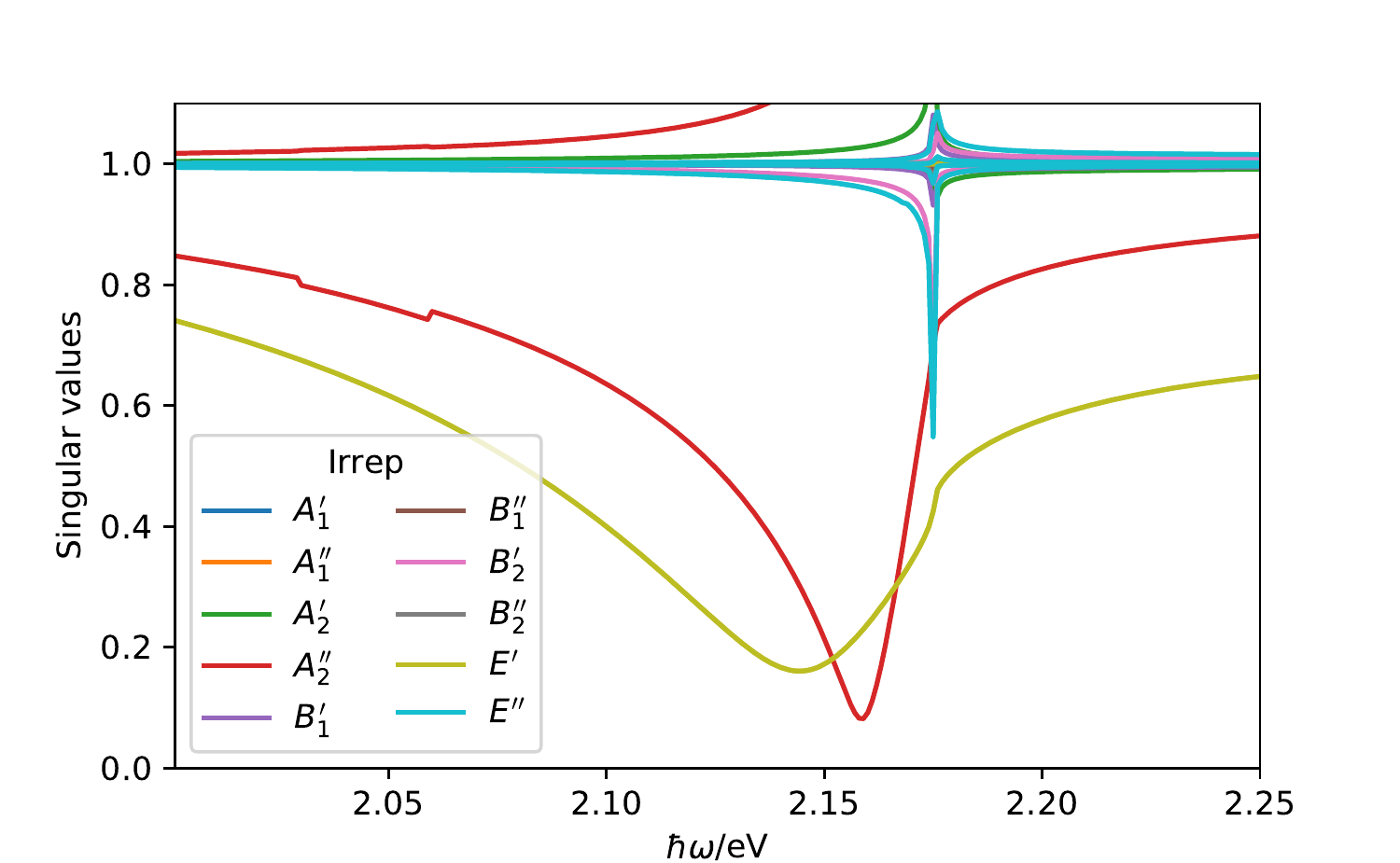}
\par\end{centering}
\caption{Singular values of the mode problem matrix $\protect\truncated{M\left(\omega,\protect\vect k=0\right)}3$
for a real frequency interval. The irreducible representations of
$D_{4h}$ are labeled with different colors. The density of the data
points on the horizontal axis is $1/\mathrm{meV}$. }\label{fig:square lattice real interval SVD}

\end{figure}

Next, we study the lattice mode problem of the same square arrays.
First we consider the mode problem exactly at the $\Gamma$ point,
$\vect k=0$. Before proceeding with more sophisticated methods, it
is often helpful to look at the singular values of mode problem matrix
$M\left(\omega,\vect k\right)$ from the lattice mode equation (\ref{eq:lattice mode equation}),
as shown in Fig. \ref{fig:square lattice real interval SVD}. This
can be always done, even with tabulated/interpolated material properties
and/or $T$-matrices. An additional insight, especially in the high-symmetry
points of the Brillouin zone, is provided by decomposition of the
matrix into irreps -- in this case of group $D_{4h}$, which corresponds
to the point group symmetry of the array at the $\Gamma$ point. Although
on the picture none of the SVDs hits manifestly zero, we see two prominent
dips in the $E'$ and $A_{2}''$ irrep subspaces, which is a sign
of an actual solution nearby in the complex plane. Moreover, there
might be some less obvious minima in the very vicinity of the diffracted
order crossing which do not appear in the picture due to rough frequency
sampling.

\begin{figure}
\begin{centering}
\includegraphics[width=0.8\columnwidth]{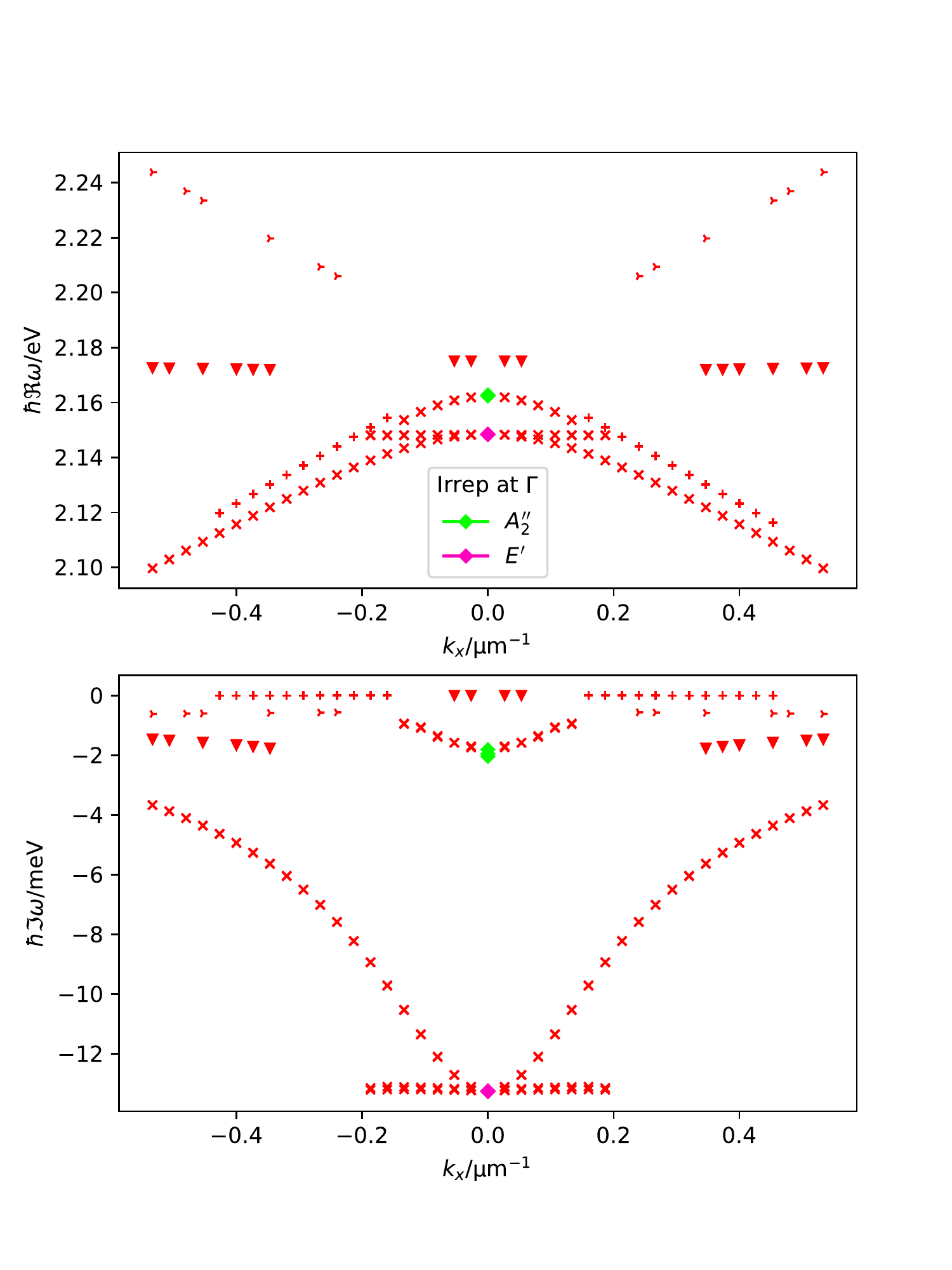}
\par\end{centering}
\caption{Solutions of the lattice mode problem $\protect\truncated{M\left(\omega,\protect\vect k\right)}3$
found using the Beyn's method near the first diffracted order crossing
at the $\Gamma$ point for $k_{y}=0$. At the $\Gamma$ point, they
are classified according to the irreducible representations of $D_{4h}$.
}\label{fig:square lattice beyn dispersion}
\end{figure}

As we have used only analytical ingredients in $M\left(\omega,\vect k\right)$,
the matrix is itself analytical, hence Beyn's algorithm can be used
to search for complex mode frequencies, which is shown in Figure \ref{fig:square lattice beyn dispersion}.
The number of the frequency point found is largely dependent on the
parameters used in Beyn's algorithm, mostly the integration contour
in the frequency space. Here we used ellipses discretised by 250 points
each, with edges nearly touching the empty lattice diffracted orders
(from either above or below in the real part), and with major axis
covering 1/5 of the interval between two diffracted orders. The residual
threshold was set to 0.1. At the $\Gamma$ point, the algorithm finds
the actual complex positions of the suspected $E'$ and $A_{2}''$
modes without a problem, as well as their continuations to the other
nearby values of $\vect k$. However, for further $\vect k$ it might
``lose track'', especially as the modes cross the diffracted orders.
As a result, the parameters of Beyn's algorithm often require manual
tuning based on the observed behaviour. 

\subsection{Effects of multipole cutoff}

In order to demonstrate some of the consequences of multipole cutoff,
we consider a square lattice with periodicity $p_{x}=p_{y}=580\,\mathrm{nm}$
filled with spherical golden nanoparticles (with Drude-Lorentz model
for permittivity; one sphere per unit cell) embedded in a medium with
a constant refractive index $n=1.52$. We vary the multipole cutoff
$l_{\max}=1,\dots,5$ and the particle radius $r=50\,\mathrm{nm},\dots,300\,\mathrm{nm}$
(note that right end of this interval is unphysical, as the spheres
touch at $r=290\,\mathrm{nm}$) We look at the lattice modes at the
$\Gamma$ point right below the diffracted order crossing at 1.406
eV using Beyn's algorithm; the integration contour for Beyn's algorithm
being a circle with centre at $\omega=\left(1.335+0i\right)\mathrm{eV}/\hbar$
and radius $70.3\,\mathrm{meV}/\hbar$, and 410 sample points. We
classify each of the found modes as one of the ten irreducible representations
of the corresponding little group at the $\Gamma$ point, $D_{4h}$.

The real and imaginary parts of the obtained mode frequencies are
shown in Fig. \ref{square lattice var lMax, r at gamma point Au}.
The most obvious (and expected) effect of the cutoff is the reduction
of the number of modes found: the case $l_{\max}=1$ (dipole-dipole
approximation) contains only the modes with nontrivial dipole excitations
($x,y$ dipoles in $\mathrm{E}'$ and $z$ dipole in $\mathrm{A_{2}''})$.
For relatively small particle sizes, the main effect of increasing
$l_{\max}$ is making the higher multipolar modes accessible at all.
As the particle radius increases, there start to appear more non-negligible
elements in the $T$-matrix, and the cutoff then affects the mode
frequencies as well.

Another effect related to mode finding is, that increasing $l_{\max}$
leads to overall decrease of the lowest singular values of the mode
problem matrix $M\left(\omega,\vect k\right)$, so that they are very
close to zero for a large frequency area, making it harder to determine
the exact roots of the mode equation (\ref{eq:lattice mode equation}),
which might lead to some spurious results: Fig. \ref{square lattice var lMax, r at gamma point Au}
shows modes with positive imaginary frequencies for $l_{\max}\ge3$,
which is unphysical (positive imaginary frequency means effective
losses of the medium, which, together with the lossy particles, prevent
emergence of propagating modes). However, the spurious frequencies
can be made disappear by tuning the parameters of Beyn's algorithm
(namely, stricter residual threshold), but that might lead to losing
legitimate results as well, especially if they are close to the integration
contour. In such cases, it is often helpful to run Beyn's algorithm
several times with different contours enclosing smaller frequency
areas.

\begin{figure}
\centering{}\includegraphics[width=1\textwidth]{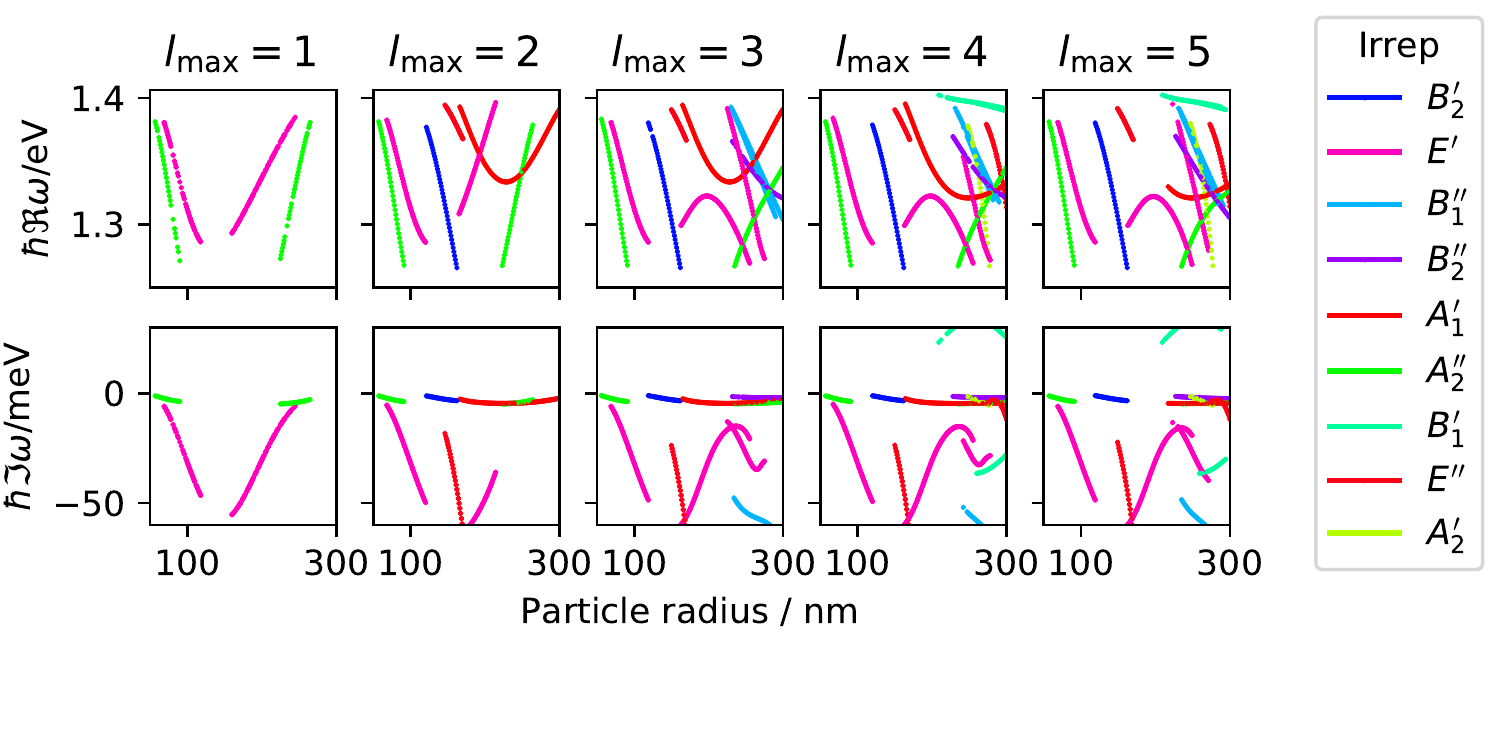}\caption{Consequences of multipole degree cutoff: Eigenfrequencies found with
the Beyn's algorithm for an infinite square lattice of golden spherical
nanoparticles with varying particle size.}\label{square lattice var lMax, r at gamma point Au}
\end{figure}

\section{Summary}

We presented two major enhancements of the electromagnetic multiple-scattering
$T$-matrix method: 1) Employing Ewald summation techniques enables
very efficient computation of lattice modes and optical response of
infinite periodic nanoparticle structures. 2) Exploiting possible
symmetries of the system by transformation into symmetry adapted basis
reduces the requirements on computational resources considerably,
enabling simulations of finite systems with tens of thousands of scatterers.
These enhancements are included into the QPMS software suite, which
we hereby make publicly available under the GNU General Public License.

\section{Acknowledgments}

We thank Nicki Källman, Kristian Arjas, Javier Cuerda and Vadim Zakomirnyi
for useful discussions. This work was supported by the Academy of
Finland under project numbers 303351, 307419, 327293, 318987 (QuantERA
project RouTe), and by the European Research Council (ERC-2013-AdG-340748-CODE).
We acknowledge the computational resources provided by the Aalto Science-IT
project.

\bibliographystyle{amsplain}
\bibliography{Tmatrix}

\selectlanguage{finnish}%

\section{Supplementary material: Derivation of the 1D and 2D lattice sums
for the 3D Helmholtz equation with general lattice offset}

\global\long\def\dual#1{\left.#1'\right.}%

\subsection{Periodic Green's functions vs. VSWF lattice sums}

\selectlanguage{english}%

\selectlanguage{finnish}%

\subsubsection{Some definitions and useful relations}

\[
\mathcal{H}_{l}^{m}\left(\vect d\right)\equiv h_{l}^{+}\left(\left|\vect d\right|\right)\ush lm\left(\uvec d\right),
\]
\[
\mathcal{J}_{l}^{m}\left(\vect d\right)\equiv j_{l}\left(\left|\vect d\right|\right)\ush lm\left(\uvec d\right).
\]
Dual spherical harmonics and waves:

\[
\int\ush lm\ushD{l'}{m'}\,\ud\Omega=\delta_{l,l'}\delta_{m,m'},
\]
\[
\dual{\mathcal{J}}{}_{l}^{m}\left(\vect d\right)\equiv j_{l}\left(\left|\vect d\right|\right)\ushD lm\left(\uvec d\right).
\]
Expansion of a plane wave: 

\[
e^{i\kappa\vect r\cdot\uvec r'}=4\pi\sum_{l,m}i^{n}\dual{\mathcal{J}}{}_{l}^{m}\left(\kappa\vect r\right)\ush lm\left(\uvec r'\right)=4\pi\sum_{l,m}i^{n}\mathcal{J}{}_{l}^{m}\left(\kappa\vect r\right)\ushD lm\left(\uvec r'\right).
\]
This one is also convention independent (similarly for $\mathcal{H}_{l}^{m}$):
\[
\mathcal{J}_{l}^{m}\left(-\vect r\right)=\left(-1\right)^{l}\mathcal{J}_{l}^{m}\left(\vect r\right).
\]

\subsubsection{Helmholtz equation and Green's functions (in 3D)}

Note that the notation does not follow Linton's (where the wavenumbers
are often implicit)

\[
\left(\nabla^{2}+\kappa^{2}\right)G^{(\kappa)}\left(\vect x,\vect x_{0}\right)=\delta\left(\vect x-\vect x_{0}\right),
\]
\begin{align*}
G_{0}^{(\kappa)}\left(\vect x,\vect x_{0}\right) & =G_{0}^{(\kappa)}\left(\vect x-\vect x_{0}\right)\\
 & =-\frac{e^{i\kappa\left|\vect x-\vect x_{0}\right|}}{4\pi\left|\vect x-\vect x_{0}\right|}\\
 & =-\frac{i\kappa}{4\pi}h_{0}^{+}\left(\kappa\left|\vect x-\vect x_{0}\right|\right)\\
 & =-\frac{i\kappa}{\sqrt{4\pi}}\mathcal{H}_{0}^{0}\left(\kappa\left|\vect x-\vect x_{0}\right|\right).
\end{align*}
In case of wacky conventions, $G_{0}^{(\kappa)}\left(\vect x,\vect x_{0}\right)=-\frac{i\kappa}{\ush 00}\mathcal{H}_{0}^{0}\left(\kappa\left|\vect x-\vect x_{0}\right|\right)$.

Lattice GF \cite[(2.3)]{linton_lattice_2010}:
\begin{equation}
G_{\Lambda}^{(\kappa)}\left(\vect s,\vect k\right)\equiv\sum_{\vect R\in\Lambda}G_{0}^{\kappa}\left(\vect s-\vect R\right)e^{i\vect k\cdot\vect R}\label{eq:Lattice GF}
\end{equation}

\subsubsection{GF expansion and lattice sum definition}

Let's define 
\[
\sigma_{l}^{m}\left(\vect s,\vect k\right)=\sum_{\vect R\in\Lambda}\mathcal{H}_{l}^{m}\left(\kappa\left(\vect s+\vect R\right)\right)e^{i\vect k\cdot\vect R},
\]
and also its dual version
\[
\dual{\sigma}{}_{l}^{m}\left(\vect s,\vect k\right)=\sum_{\vect R\in\Lambda}\dual{\mathcal{H}}{}_{l}^{m}\left(\kappa\left(\vect s+\vect R\right)\right)e^{i\vect k\cdot\vect R}.
\]

Inspired by \cite[(4.1)]{linton_lattice_2010}: assuming that $\vect s\notin\Lambda$,
let's expand the lattice Green's function around $\vect s$:

\[
G_{\Lambda}^{(\kappa)}\left(\vect s+\vect r,\vect k\right)=-i\kappa\sum_{l,m}\tau_{l}^{m}\left(\vect s,\vect k\right)\mathcal{J}_{l}^{m}\left(\kappa\vect r\right),
\]
and multiply with a dual SH + integrate
\begin{align}
\int\ud\Omega_{\vect r}\,G_{\Lambda}^{(\kappa)}\left(\vect s+\vect r,\vect k\right)\ushD{l'}{m'}\left(\uvec r\right) & =-i\kappa\sum_{l,m}\tau_{l}^{m}\left(\vect s,\vect k\right)j_{l}\left(\kappa\left|\vect r\right|\right)\delta_{ll'}\delta_{mm'}\nonumber \\
 & =-i\kappa\tau_{l'}^{m'}\left(\vect s,\vect k\right)j_{l'}\left(\kappa\left|\vect r\right|\right).\label{eq:tau extraction}
\end{align}
The expansion coefficients $\tau_{l}^{m}\left(\vect s,\vect k\right)$
is then typically extracted by taking the limit $\left|\vect r\right|\to0$.

The relation between $\sigma_{l}^{m}\left(\vect s,\vect k\right)$
and $\tau_{l}^{m}\left(\vect s,\vect k\right)$ can be obtained e.g.
from the addition theorem for scalar spherical wavefunctions \cite[(C.3)]{linton_lattice_2010},
\[
\mathcal{H}_{l}^{m}\left(\vect a+\vect b\right)=\sum_{l'm'}S_{ll'}^{mm'}\left(\vect b\right)\mathcal{J}_{l'}^{m'}\left(\vect a\right),\quad\left|\vect a\right|<\left|\vect b\right|,
\]
where for the zeroth degree and order one has \cite[(C.3)]{linton_lattice_2010}\footnote{In a totally convention-independent version probably looks like $S_{0l'}^{0m'}\left(\vect b\right)=\ush 00\dual{\mathcal{H}}{}_{l'}^{m'}\left(-\vect b\right)$,
but the $Y_{0}^{0}$ will cancel with the expression for GF anyways,
so no harm to the final result.}
\[
S_{0l'}^{0m'}\left(\vect b\right)=\sqrt{4\pi}\dual{\mathcal{H}}{}_{l'}^{m'}\left(-\vect b\right).
\]
From the lattice GF definition (\ref{eq:Lattice GF})
\begin{align*}
G_{\Lambda}^{(\kappa)}\left(\vect s+\vect r,\vect k\right) & \equiv\frac{-i\kappa}{\sqrt{4\pi}}\sum_{\vect R\in\Lambda}\mathcal{H}_{0}^{0}\left(\kappa\left(\vect s+\vect r-\vect R\right)\right)e^{i\vect k\cdot\vect R}\\
 & =\frac{-i\kappa}{\sqrt{4\pi}}\sum_{\vect R\in\Lambda}\mathcal{H}_{0}^{0}\left(\kappa\left(\vect s+\vect r-\vect R\right)\right)e^{i\vect k\cdot\vect R}\\
 & =\frac{-i\kappa}{\sqrt{4\pi}}\sum_{\vect R\in\Lambda}\sum_{l'm'}S_{0l'}^{0m'}\left(\kappa\left(\vect s-\vect R\right)\right)\mathcal{J}_{l'}^{m'}\left(\kappa\vect r\right)e^{i\vect k\cdot\vect R}\\
 & =-i\kappa\sum_{\vect R\in\Lambda}\sum_{lm}\dual{\mathcal{H}}{}_{l}^{m}\left(-\kappa\left(\vect s-\vect R\right)\right)\mathcal{J}_{l}^{m}\left(\kappa\vect r\right)e^{i\vect k\cdot\vect R},
\end{align*}
and mutliplying with a dual SH and integrating 
\begin{align*}
\int\ud\Omega_{\vect r}\,G_{\Lambda}^{(\kappa)}\left(\vect s+\vect r,\vect k\right)\ushD{l'}{m'}\left(\uvec r\right) & =-i\kappa\sum_{\vect R\in\Lambda}\sum_{lm}\dual{\mathcal{H}}{}_{l}^{m}\left(-\kappa\left(\vect s-\vect R\right)\right)j_{l}\left(\kappa\left|\vect r\right|\right)\delta_{ll'}\delta_{mm'}e^{i\vect k\cdot\vect R}\\
 & =-i\kappa\sum_{\vect R\in\Lambda}\dual{\mathcal{H}}{}_{l'}^{m'}\left(\kappa\left(-\vect s+\vect R\right)\right)j_{l'}\left(\kappa\left|\vect r\right|\right)e^{i\vect k\cdot\vect R}\\
 & =-i\kappa\dual{\sigma}{}_{l'}^{m'}\left(-\vect s,\vect k\right)j_{l'}\left(\kappa\left|\vect r\right|\right),
\end{align*}
and comparing with (\ref{eq:tau extraction}) we have
\[
\tau_{l}^{m}\left(\vect s,\vect k\right)=\dual{\sigma}{}_{l}^{m}\left(-\vect s,\vect k\right).
\]

\subsection{Derivation of the 1D and 2D lattice sum}

With \cite{linton_lattice_2010} in hand, the short-range part is
rather easy. Let's get the long-range part.

We first need to find the long-range part of the expansion coefficient

\begin{equation}
\tau_{l'}^{m'}\left(\vect s,\vect k\right)=\frac{i}{\kappa j_{l'}\left(\kappa\left|\vect r\right|\right)}\int\ud\Omega_{\vect r}\,G_{\Lambda}^{(\kappa)}\left(\vect s+\vect r,\vect k\right)\ushD{l'}{m'}\left(\uvec r\right).\label{eq:tau extraction formula}
\end{equation}

We take \cite[(2.24)]{linton_lattice_2010} with slightly modified
notation $\left(\vect k_{\vect K}\equiv\vect K+\vect k\right)$
\[
G_{\Lambda}^{(1;\kappa)}\left(\vect r\right)=-\frac{1}{2\pi^{d_{c}/2}\mathcal{A}}\sum_{\vect K\in\Lambda^{*}}e^{i\vect k_{\vect K}\cdot\vect r}\int_{1/\eta}^{\infty e^{i\pi/4}}e^{-\kappa^{2}\gamma^{2}t^{2}/4}e^{-\left|\vect r^{\bot}\right|^{2}/t^{2}}t^{1-d_{c}}\ud t
\]
or, evaluated at point $\vect s+\vect r$ instead
\[
G_{\Lambda}^{(1;\kappa)}\left(\vect s+\vect r\right)=-\frac{1}{2\pi^{d_{c}/2}\mathcal{A}}\sum_{\vect K\in\Lambda^{*}}e^{i\vect k_{\vect K}\cdot\left(\vect s+\vect r\right)}\int_{1/\eta}^{\infty e^{i\pi/4}}e^{-\kappa^{2}\gamma^{2}t^{2}/4}e^{-\left|\vect s^{\bot}+\vect r^{\bot}\right|^{2}/t^{2}}t^{1-d_{c}}\ud t.
\]
The integral can be by substitutions taken into the form 
\[
G_{\Lambda}^{(1;\kappa)}\left(\vect s+\vect r\right)=-\frac{1}{2\pi^{d_{c}/2}\mathcal{A}}\sum_{\vect K\in\Lambda^{*}}e^{i\vect k_{\vect K}\cdot\left(\vect s+\vect r\right)}\int_{\kappa^{2}\gamma_{m}^{2}/4\eta^{2}}^{\infty\exp\left(i\pi/2\right)}e^{-\tau}e^{-\left|\vect s_{\bot}+\vect r_{\bot}\right|^{2}\kappa^{2}\gamma_{m}^{2}/4\tau}\tau^{-\frac{d_{c}}{2}}\ud\tau.
\]

Let's do the integration to get $\tau_{l}^{m}\left(\vect s,\vect k\right)$
\begin{multline*}
\int\ud\Omega_{\vect r}\,G_{\Lambda}^{(1;\kappa)}\left(\vect s+\vect r\right)\ushD{l'}{m'}\left(\uvec r\right)=-\frac{1}{2\pi^{d_{c}/2}\mathcal{A}}\int\ud\Omega_{\vect r}\,\ushD{l'}{m'}\left(\uvec r\right)\sum_{\vect K\in\Lambda^{*}}e^{i\vect k_{\vect K}\cdot\left(\vect s+\vect r\right)}\times\\
\times\int_{\kappa^{2}\gamma_{\vect k_{\vect K}}^{2}/4\eta^{2}}^{\infty\exp\left(i\pi/2\right)}e^{-\tau}e^{-\left|\vect s_{\bot}+\vect r_{\bot}\right|^{2}\kappa^{2}\gamma_{\vect k_{\vect K}}^{2}/4\tau}\tau^{-\frac{d_{c}}{2}}\ud\tau.
\end{multline*}
The $\vect r$-dependent plane wave factor can be also written as
\begin{align*}
e^{i\vect k_{\vect K}\cdot\vect r} & =e^{i\left|\vect k_{\vect K}\right|\vect r\cdot\uvec{\vect k_{\vect K}}}=4\pi\sum_{lm}i^{l}\dual{\mathcal{J}}{}_{l}^{m}\left(\left|\vect k_{\vect K}\right|\vect r\right)\ush lm\left(\uvec{\vect k_{\vect K}}\right)\\
 & =4\pi\sum_{lm}i^{l}j_{l}\left(\left|\vect k_{\vect K}\right|\left|\vect r\right|\right)\ushD lm\left(\uvec{\vect r}\right)\ush lm\left(\uvec{\vect k_{\vect K}}\right),
\end{align*}
so
\begin{multline*}
\int\ud\Omega_{\vect r}\,G_{\Lambda}^{(1;\kappa)}\left(\vect s+\vect r\right)\ushD{l'}{m'}\left(\uvec r\right)=-\frac{1}{2\pi^{d_{c}/2}\mathcal{A}}\int\ud\Omega_{\vect r}\,\ushD{l'}{m'}\left(\uvec r\right)\frac{1}{2\pi\mathcal{A}}\times\\
\times\sum_{\vect K\in\Lambda^{*}}e^{i\vect k_{\vect K}\cdot\vect s}\sum_{lm}4\pi i^{l}j_{l}\left(\left|\vect k_{\vect K}\right|\left|\vect r\right|\right)\ushD lm\left(\uvec r\right)\ush lm\left(\uvec{\vect k_{\vect K}}\right)\times\\
\times\int_{\kappa^{2}\gamma_{\vect{\vect k_{\vect K}}}^{2}/4\eta^{2}}^{\infty\exp\left(i\pi/2\right)}e^{-\tau}e^{-\left|\vect s_{\bot}+\vect r_{\bot}\right|^{2}\kappa^{2}\gamma_{\vect{\vect k_{\vect K}}}^{2}/4\tau}\tau^{-\frac{d_{c}}{2}}\ud\tau.
\end{multline*}

We also have
\begin{align*}
e^{-\left|\vect s_{\bot}+\vect r_{\bot}\right|^{2}\kappa^{2}\gamma_{\vect K}^{2}/4\tau} & =e^{-\left(\left|\vect s_{\bot}\right|^{2}+\left|\vect r_{\bot}\right|^{2}+2\vect r_{\bot}\cdot\vect s_{\bot}\right)\kappa^{2}\gamma_{\vect K}^{2}/4\tau}\\
 & =e^{-\left|\vect s_{\bot}\right|^{2}\kappa^{2}\gamma_{\vect K}^{2}/4\tau}\sum_{j=0}^{\infty}\frac{1}{j!}\left(-\frac{\left(\left|\vect r_{\bot}\right|^{2}+2\vect r_{\bot}\cdot\vect s_{\bot}\right)\kappa^{2}\gamma_{\vect K}^{2}}{4\tau}\right)^{j},
\end{align*}
hence
\begin{multline*}
\int\ud\Omega_{\vect r}\,G_{\Lambda}^{(1;\kappa)}\left(\vect s+\vect r\right)\ushD{l'}{m'}\left(\uvec r\right)=-\frac{1}{2\pi^{d_{c}/2}\mathcal{A}}\int\ud\Omega_{\vect r}\,\ushD{l'}{m'}\left(\uvec r\right)\sum_{\vect K\in\Lambda^{*}}e^{i\vect k_{\vect K}\cdot\vect s}\times\\
\times\sum_{lm}4\pi i^{l}j_{l}\left(\left|\vect k_{\vect K}\right|\left|\vect r\right|\right)\ushD lm\left(\uvec r\right)\ush lm\left(\uvec{\vect k_{\vect K}}\right)\times\\
\quad\times\sum_{j=0}^{\infty}\frac{1}{j!}\left(-\frac{\left(\left|\vect r_{\bot}\right|^{2}+2\vect r_{\bot}\cdot\vect s_{\bot}\right)\kappa^{2}\gamma_{\vect{\vect k_{\vect K}}}^{2}}{4}\right)^{j}\times\\
\times\underbrace{\int_{\kappa^{2}\gamma_{\vect K}^{2}/4\eta^{2}}^{\infty\exp\left(i\pi/2\right)}e^{-\tau}e^{-\left|\vect s_{\bot}\right|^{2}\kappa^{2}\gamma_{\vect K}^{2}/4\tau}\tau^{-\frac{d_{c}}{2}-j}\ud\tau}_{\Delta_{j}^{\left(d_{\Lambda}\right)}}\\
=-\frac{1}{2\pi^{d_{c}/2}\mathcal{A}}\sum_{\vect K\in\Lambda^{*}}e^{i\vect k_{\vect K}\cdot\vect s}\sum_{lm}4\pi i^{l}j_{l}\left(\left|\vect k_{\vect K}\right|\left|\vect r\right|\right)\ush lm\left(\uvec{\vect k_{\vect K}}\right)\sum_{j=0}^{\infty}\frac{\Delta_{j}^{\left(d_{\Lambda}\right)}}{j!}\times\\
\quad\times\int\ud\Omega_{\vect r}\,\ushD{l'}{m'}\left(\uvec r\right)\ushD lm\left(\uvec r\right)\left(-\frac{\left(\left|\vect r_{\bot}\right|^{2}+2\vect r_{\bot}\cdot\vect s_{\bot}\right)\kappa^{2}\gamma_{\vect k_{\vect K}}^{2}}{4}\right)^{j}\\
=-\frac{1}{2\pi^{d_{c}/2}\mathcal{A}}\sum_{\vect K\in\Lambda^{*}}e^{i\vect k_{\vect K}\cdot\vect s}\sum_{lm}4\pi i^{l}j_{l}\left(\left|\vect k_{\vect K}\right|\left|\vect r\right|\right)\ush lm\left(\uvec{\vect k_{\vect K}}\right)\sum_{j=0}^{\infty}\frac{\left(-1\right)^{j}}{j!}\Delta_{j}^{\left(d_{\Lambda}\right)}\times\\
\quad\times\left(\frac{\kappa\gamma_{\vect{\vect k_{\vect K}}}}{2}\right)^{2j}\sum_{k=0}^{j}\int\ud\Omega_{\vect r}\,\ushD{l'}{m'}\left(\uvec r\right)\ushD lm\left(\uvec r\right)\left|\vect r_{\bot}\right|^{2(j-k)}\left(2\vect r_{\bot}\cdot\vect s_{\bot}\right)^{k}.
\end{multline*}
The integral $\Delta_{j}^{\left(d_{\Lambda}\right)}$ is (for the
2D case) equivalent to that in \cite{kambe_theory_1968}.

If we label $\left|\vect r_{\bot}\right|\left|\vect s_{\bot}\right|\cos\varphi\equiv\vect r_{\bot}\cdot\vect s_{\bot}$,
we have
\begin{multline*}
\int\ud\Omega_{\vect r}\,G_{\Lambda}^{(1;\kappa)}\left(\vect s+\vect r\right)\ushD{l'}{m'}\left(\uvec r\right)=-\frac{1}{2\pi^{d_{c}/2}\mathcal{A}}\sum_{\vect K\in\Lambda^{*}}e^{i\vect k_{\vect K}\cdot\vect s}\sum_{lm}4\pi i^{l}j_{l}\left(\left|\vect k_{\vect K}\right|\left|\vect r\right|\right)\ush lm\left(\uvec{\vect k_{\vect K}}\right)\times\\
\times\sum_{j=0}^{\infty}\frac{\left(-1\right)^{j}}{j!}\Delta_{j}^{\left(d_{\Lambda}\right)}\left(\frac{\kappa\gamma_{\vect k_{\vect K}}}{2}\right)^{2j}\sum_{k=0}^{j}\left(2\left|\vect s_{\bot}\right|\right)^{k}\int\ud\Omega_{\vect r}\,\ushD{l'}{m'}\left(\uvec r\right)\ushD lm\left(\uvec r\right)\left|\vect r_{\bot}\right|^{2j-k}\left(\cos\varphi\right)^{k},
\end{multline*}
and if we label $\left|\vect r\right|\sin\vartheta\equiv\left|\vect r_{\bot}\right|$
\begin{multline*}
\int\ud\Omega_{\vect r}\,G_{\Lambda}^{(1;\kappa)}\left(\vect s+\vect r\right)\ushD{l'}{m'}\left(\uvec r\right)=-\frac{1}{2\pi^{d_{c}/2}\mathcal{A}}\sum_{\vect K\in\Lambda^{*}}e^{i\vect k_{\vect K}\cdot\vect s}\times\\
\times\sum_{lm}4\pi i^{l}j_{l}\left(\left|\vect k_{\vect K}\right|\left|\vect r\right|\right)\ush lm\left(\uvec{\vect k_{\vect K}}\right)\sum_{j=0}^{\infty}\frac{\left(-1\right)^{j}}{j!}\Delta_{j}^{\left(d_{\Lambda}\right)}\left(\frac{\kappa\gamma_{\vect k_{\vect K}}}{2}\right)^{2j}\times\\
\times\sum_{k=0}^{j}\left|\vect r\right|^{2j-k}\left(2\left|\vect s_{\bot}\right|\right)^{k}\int\ud\Omega_{\vect r}\,\ushD{l'}{m'}\left(\uvec r\right)\ushD lm\left(\uvec r\right)\left(\sin\vartheta\right)^{2j-k}\left(\cos\varphi\right)^{k}.
\end{multline*}
Now let's put the RHS into (\ref{eq:tau extraction formula}) and
try eliminating some sum by taking the limit $\left|\vect r\right|\to0$.
We have $j_{l}\left(\left|\vect k_{\vect K}\right|\left|\vect r\right|\right)\sim\left(\left|\vect k_{\vect K}\right|\left|\vect r\right|\right)^{l}/\left(2l+1\right)!!$;
the denominator from (\ref{eq:tau extraction formula}) behaves like
$j_{l'}\left(\kappa\left|\vect r\right|\right)\sim\left(\kappa\left|\vect r\right|\right)^{l'}/\left(2l'+1\right)!!.$
The leading terms are hence those with $\left|\vect r\right|^{l-l'+2j-k}$.
So 
\begin{multline*}
\tau_{l'}^{m'}\left(\vect s,\vect k\right)=\frac{-i}{2\pi^{d_{c}/2}\mathcal{A}\kappa^{1+l'}}\left(2l'+1\right)!!\sum_{\vect K\in\Lambda^{*}}e^{i\vect k_{\vect K}\cdot\vect s}\sum_{lm}4\pi i^{l}\frac{\left|\vect k_{\vect K}\right|^{l}}{\left(2l+1\right)!!}\ush lm\left(\uvec{\vect k_{\vect K}}\right)\times\\
\times\sum_{j=0}^{\infty}\frac{\left(-1\right)^{j}}{j!}\Delta_{j}^{\left(d_{\Lambda}\right)}\left(\frac{\kappa\gamma_{\vect k_{\vect K}}}{2}\right)^{2j}\sum_{k=0}^{j}\delta_{l'-l,2j-k}\left(2\left|\vect s_{\bot}\right|\right)^{k}\times\\
\times\int\ud\Omega_{\vect r}\,\ushD{l'}{m'}\left(\uvec r\right)\ushD lm\left(\uvec r\right)\left(\sin\vartheta\right)^{l'-l}\left(\cos\varphi\right)^{k}.
\end{multline*}
Let's now focus on rearranging the sums; we have
\[
S(l')\equiv\sum_{l=0}^{\infty}\sum_{j=0}^{\infty}\sum_{k=0}^{j}\delta_{l'-l,2j-k}f(l',l,j,k)=\sum_{l=0}^{\infty}\sum_{j=0}^{\infty}\sum_{k=0}^{j}\delta_{l'-l,2j-k}f(l',l,j,2j-l'+l).
\]
We have $0\le k\le j$, hence $0\le2j-l'+l\le j$, hence $-2j\le-l'+l\le-j$,
hence also $l'-2j\le l\le l'-j$, which gives the opportunity to swap
the $l,j$ sums and the $l$-sum becomes finite; so also consuming
$\sum_{k=0}^{j}\delta_{l'-l,2j-k}$ we get 
\[
S(l')=\sum_{j=0}^{\infty}\sum_{l=\max(0,l'-2j)}^{l'-j}f(l',l,j,2j-l'+l).
\]
Finally, we see that the interval of valid $l$ becomes empty when
$l'-j<0$, i.e. $j>l'$; so we get a finite sum
\[
S(l')=\sum_{j=0}^{l'}\sum_{l=\max(0,l'-2j)}^{l'-j}f(l',l,j,2j-l'+l).
\]
Applying rearrangement,
\begin{multline*}
\tau_{l'}^{m'}\left(\vect s,\vect k\right)=\frac{-i}{2\pi^{d_{c}/2}\mathcal{A}\kappa}\frac{\left(2l'+1\right)!!}{\kappa^{l'}}\sum_{\vect K\in\Lambda^{*}}e^{i\vect k_{\vect K}\cdot\vect s}\sum_{j=0}^{l'}\frac{\left(-1\right)^{j}}{j!}\Delta_{j}^{\left(d_{\Lambda}\right)}\left(\frac{\kappa\gamma_{\vect k_{\vect K}}}{2}\right)^{2j}\times\\
\times\sum_{l=\max\left(0,l'-2j\right)}^{l'-j}4\pi i^{l}\left(2\left|\vect s_{\bot}\right|\right)^{2j-l'+l}\frac{\left|\vect k_{\vect K}\right|^{l}}{\left(2l+1\right)!!}\times\\
\times\sum_{m=-l}^{l}\ush lm\left(\uvec{\vect k_{\vect K}}\right)\int\ud\Omega_{\vect r}\,\ushD{l'}{m'}\left(\uvec r\right)\ushD lm\left(\uvec r\right)\left(\sin\vartheta\right)^{l'-l}\left(\cos\varphi\right)^{2j-l'+l},
\end{multline*}
or replacing the angles with their original definition,
\begin{multline*}
\tau_{l'}^{m'}\left(\vect s,\vect k\right)=\frac{-i}{2\pi^{d_{c}/2}\mathcal{A}\kappa}\frac{\left(2l'+1\right)!!}{\kappa^{l'}}\sum_{\vect K\in\Lambda^{*}}e^{i\vect k_{\vect K}\cdot\vect s}\sum_{j=0}^{l'}\frac{\left(-1\right)^{j}}{j!}\Delta_{j}^{\left(d_{\Lambda}\right)}\left(\frac{\kappa\gamma_{\vect K}}{2}\right)^{2j}\times\\
\times\sum_{l=\max\left(0,l'-2j\right)}^{l'-j}4\pi i^{l}\left(2\left|\vect s_{\bot}\right|\right)^{2j-l'+l}\frac{\left|\vect k_{\vect K}\right|^{l}}{\left(2l+1\right)!!}\\
\times\sum_{m=-l}^{l}\ush lm\left(\uvec K\right)\int\ud\Omega_{\vect r}\,\ushD{l'}{m'}\left(\uvec r\right)\ushD lm\left(\uvec r\right)\left(\frac{\left|\vect r_{\bot}\right|}{\left|\vect r\right|}\right)^{l'-l}\left(\frac{\vect r_{\bot}\cdot\vect s_{\bot}}{\left|\vect r_{\bot}\right|\left|\vect s_{\bot}\right|}\right)^{2j-l'+l},
\end{multline*}
and if we want a $\sigma_{l'}^{m'}\left(\vect s,\vect k\right)$ instead,
we reverse the sign of $\vect s$ and replace all spherical harmonics
with their dual counterparts:
\begin{multline*}
\sigma_{l'}^{m'}\left(\vect s,\vect k\right)=\frac{-i}{2\pi^{d_{c}/2}\mathcal{A}\kappa}\frac{\left(2l'+1\right)!!}{\kappa^{l'}}\sum_{\vect K\in\Lambda^{*}}e^{-i\vect k_{\vect K}\cdot\vect s}\sum_{j=0}^{l'}\frac{\left(-1\right)^{j}}{j!}\Delta_{j}^{\left(d_{\Lambda}\right)}\left(\frac{\kappa\gamma_{\vect k_{\vect K}}}{2}\right)^{2j}\times\\
\times\sum_{l=\max\left(0,l'-2j\right)}^{l'-j}4\pi i^{l}\left(2\left|\vect s_{\bot}\right|\right)^{2j-l'+l}\frac{\left|\vect k_{\vect K}\right|^{l}}{\left(2l+1\right)!!}\times\\
\times\sum_{m=-l}^{l}\ushD lm\left(\uvec{\vect k_{\vect K}}\right)\int\ud\Omega_{\vect r}\,\ush{l'}{m'}\left(\uvec r\right)\ush lm\left(\uvec r\right)\left(\frac{\left|\vect r_{\bot}\right|}{\left|\vect r\right|}\right)^{l'-l}\left(\frac{-\vect r_{\bot}\cdot\vect s_{\bot}}{\left|\vect r_{\bot}\right|\left|\vect s_{\bot}\right|}\right)^{2j-l'+l},
\end{multline*}
and remembering that in the plane wave expansion the ``duality''
is interchangeable,
\begin{multline*}
\sigma_{l'}^{m'}\left(\vect s,\vect k\right)=\frac{-i}{2\pi^{d_{c}/2}\mathcal{A}\kappa}\frac{\left(2l'+1\right)!!}{\kappa^{l'}}\sum_{\vect K\in\Lambda^{*}}e^{-i\vect k_{\vect K}\cdot\vect s}\sum_{j=0}^{l'}\frac{\left(-1\right)^{j}}{j!}\Delta_{j}^{\left(d_{\Lambda}\right)}\left(\frac{\kappa\gamma_{\vect k_{\vect K}}}{2}\right)^{2j}\times\\
\times\sum_{l=\max\left(0,l'-2j\right)}^{l'-j}4\pi i^{l}\left(2\left|\vect s_{\bot}\right|\right)^{2j-l'+l}\frac{\left|\vect k_{\vect K}\right|^{l}}{\left(2l+1\right)!!}\times\\
\times\sum_{m=-l}^{l}\ush lm\left(\uvec{\vect k_{\vect K}}\right)\underbrace{\int\ud\Omega_{\vect r}\,\ush{l'}{m'}\left(\uvec r\right)\ushD lm\left(\uvec r\right)\left(\frac{\left|\vect r_{\bot}\right|}{\left|\vect r\right|}\right)^{l'-l}\left(\frac{-\vect r_{\bot}\cdot\vect s_{\bot}}{\left|\vect r_{\bot}\right|\left|\vect s_{\bot}\right|}\right)^{2j-l'+l}}_{\equiv A_{l',l,m',m,j}^{\left(d_{\Lambda}\right)}}.
\end{multline*}
The angular integral is easier to evaluate when $d_{\Lambda}=2$,
because then $\vect r_{\bot}$ is parallel (or antiparallel) to $\vect s_{\bot}$,
which gives 
\[
A_{l',l,m',m,j}^{\left(2\right)}=\left(-\frac{\vect r_{\bot}\cdot\vect s_{\bot}}{\left|\vect r_{\bot}\cdot\vect s_{\bot}\right|}\right)^{2j-l'+l}\int\ud\Omega_{\vect r}\,\ush{l'}{m'}\left(\uvec r\right)\ushD lm\left(\uvec r\right)\left(\frac{\left|\vect r_{\bot}\right|}{\left|\vect r\right|}\right)^{2j}
\]
and if we set the normal of the lattice correspond to the $z$ axis,
the azimuthal part of the integral will become zero unless $m'=m$
for any meaningful spherical harmonics convention, and the polar part
for the only nonzero case has a closed-form expression, see e.g. \cite[(A.15)]{linton_lattice_2010},
so one arrives at an expression similar to \cite[(3.15)]{kambe_theory_1968}\foreignlanguage{english}{
\begin{multline}
\sigma_{l,m}^{\left(\mathrm{L},\eta\right)}\left(\vect k,\vect s\right)=-\frac{i^{l+1}}{\kappa^{2}\mathcal{A}}\pi^{3/2}2\left(\left(l-m\right)/2\right)!\left(\left(l+m\right)/2\right)!\times\\
\times\sum_{\vect K\in\Lambda^{*}}e^{i\vect k_{\vect K}\cdot\vect s}\ush lm\left(\vect k_{\vect K}\right)\sum_{j=0}^{l-\left|m\right|}\left(-1\right)^{j}\gamma_{\vect k_{\vect K}}^{2}{}^{2j+1}\times\\
\times\Delta_{j}\left(\frac{\kappa^{2}\gamma_{\vect k_{\vect K}}^{2}}{4\eta^{2}},-i\kappa\gamma_{\vect k_{\vect K}}^{2}s_{\perp}\right)\times\\
\times\sum_{\substack{s\\
j\le s\le\min\left(2j,l-\left|m\right|\right)\\
l-j+\left|m\right|\,\mathrm{even}
}
}\frac{1}{\left(2j-s\right)!\left(s-j\right)!}\frac{\left(-\kappa s_{\perp}\right)^{2j-s}\left(\left|\vect k_{\vect K}\right|/\kappa\right)^{l-s}}{\left(\frac{1}{2}\left(l-m-s\right)\right)!\left(\frac{1}{2}\left(l+m-s\right)\right)!}\label{eq:Ewald in 3D long-range part 1D 2D-1}
\end{multline}
where $s_{\perp}\equiv\vect s\cdot\uvec z=\vect s_{\bot}\cdot\uvec z$.
If $d_{\Lambda}=1$, the angular becomes more complicated to evaluate
due to the different behaviour of the $\vect r_{\bot}\cdot\vect s_{\bot}/\left|\vect r_{\bot}\right|\left|\vect s_{\bot}\right|$
factor. The choice of coordinates can make most of the terms dissapear:
if the lattice is set parallel to the $z$ axis, $A_{l',l,m',m,j}^{\left(1\right)}$
is zero unless $m=0$, but one still has 
\[
A_{l',l,m',0,j}^{\left(1\right)}=\pi\delta_{m',l'-l-2j}\lambda'_{l0}\lambda_{l'm'}\int_{-1}^{1}\ud x\,P_{l'}^{m'}\left(x\right)P_{l}^{0}\left(x\right)\left(1-x^{2}\right)^{\frac{l'-l}{2}}
\]
where $\lambda_{lm}$ are constants depending on the conventions for
spherical harmonics. This does not seem to have such a nice closed-form
expression as in the 2D case, but it can be evaluated e.g. using the
common recurrence relations for associated Legendre polynomials. Of
course when $\vect s=0$, one gets relatively nice closed expressions,
such as those in \cite{linton_lattice_2010}.}\selectlanguage{english}%

\end{document}